\documentclass{vldb}
\usepackage{pifont}
\usepackage{verbatim}
\usepackage{times}
\usepackage{listings}
\usepackage{color}
\usepackage{balance}
\usepackage{graphicx}
\usepackage[final]{microtype}
\usepackage{url}
\usepackage{soul}
\usepackage[nocompress]{cite}
\usepackage{todonotes}
\usepackage{tabularx}
\usepackage{balance}
\usepackage[justification=centerlast, skip=12pt]{caption}
\newcommand{\minisec}[1]{\noindent\textbf{#1.}}

\newcounter{mandatoryfeature}
\newcounter{optionalfeature}
\newcommand{\manfeature}[2]{\noindent\textsf{\textit{\textbf{Feature \refstepcounter{mandatoryfeature}\themandatoryfeature\label{#2}:} #1.}}}

\newcommand{\reactdb}{\textsc{ReactDB}}

\newcommand{\checkmarkwithcross}{\checkmark\kern-1.1ex\raisebox{1ex}{\rotatebox[origin=c]{125}{--}}}
{\newcommand{\xmark}{\ding{55}}
\lstset{
    mathescape
}

\lstset{language=C++,
    basicstyle=\scriptsize,
    keywordstyle=\color{blue}\ttfamily,
    stringstyle=\color{red}\ttfamily,
    breaklines=true,
    showspaces=false,
    showstringspaces=false,
    breakatwhitespace=false,       
    frame=single,
    morekeywords={override, State, Result, Method, tuple, Relation, Durable, Actor, Actors, Create, With, In, list, get, when_all, of, type, names, future, nondurable, select, from, where, join, sum, detach, distinct, commit, exactly, once, on, grant, access, to, methods, revoke, all, drop, open, string, timestamp, encrypted, insert, into, values, group, by, having, current_time, avg, stddev, case, when, then, update, set, foreach, table, inner, over, partition, order, rows, between, current, row, following, as, desc, apply, with, args, on},
    sensitive=false
}

\vldbTitle{Actor-Relational Database Systems: A Manifesto}
\vldbAuthors{Vivek Shah and Marcos Antonio Vaz Salles}
\vldbDOI{https://doi.org/TBD}
\vldbVolume{12}
\vldbNumber{xxx}
\vldbYear{2019}

\begin{document}
\makeatletter
\def\@copyrightspace{\relax}
\def\@mkbibcitation{\relax}
\makeatother
\title{Actor-Relational Database Systems: A Manifesto}
\numberofauthors{2} 
\author{
\alignauthor
Vivek Shah\\
\affaddr{University of Copenhagen, Denmark}\\
\email{bonii@di.ku.dk}
\alignauthor
Marcos Antonio Vaz Salles\\
\affaddr{University of Copenhagen, Denmark}\\
\email{vmarcos@di.ku.dk}
}
\maketitle
\begin{abstract}
Interactive data-intensive applications are becoming ever more pervasive in domains such as finance, web applications, mobile computing, and Internet of Things. Increasingly, these applications are being deployed in sophisticated parallel and distributed hardware infrastructures. With this growing diversity of the software and hardware landscape, there is a pressure on programming models and systems to enable developers to design modular, scalable, efficient, and consistent data-intensive applications. In response to this challenge, recent research has advocated the integration of actor programming models and database management. This integration promises to help developers build logically distributed micro-applications well adapted to modern hardware trends as opposed to existing approaches targeted at optimizing monolithic applications. 

Towards this aim, in this paper we analyze, make the case for, and present a broad vision of actor-relational database systems. We argue why the time is ripe today to examine the research opportunities afforded by this emerging system paradigm. Based on this discussion, we present design principles as well as candidate feature sets to help concretize the vision for such systems. To illustrate the usefulness of the proposed feature set and motivate the need for this class of systems, we show a detailed case study inspired by a smart supermarket application with self-checkout, along with evidence for performance benefits on modern hardware.

\end{abstract}
\section{Introduction}
Online data-intensive services are becoming increasingly ubiquitous, requiring management of substantial data along with low-latency interactions. These interactive data-intensive applications include online games, social networks, financial systems, crowd collaboration and synchronization platforms, operational analytics data management, web applications, as well as upcoming Internet-of-Things (IoT) and mobile computing platforms~\cite{BohmDMPS16:Operational,Brook15:Finance,Stonebraker12:NewSQL,WhiteKGGD07:Games,GuinardTMW11:IoT, who-is-using-erlang,uses-of-akka,who-is-using-orleans}. The landscape of computing hardware is also undergoing a massive change with the advent of highly parallel hardware, fast persistent memory and networking technologies, as well as deployment architectures spanning heterogenous computing nodes over multiple data centers and interconnecting edge nodes and devices. As a result, there is an increasing pressure on systems used to build and deploy these services to deliver high performance, resource efficiency, elasticity, failure resilience and programming productivity. 

Actor frameworks and languages, such as Akka~\cite{akka-web-documentation}, Erlang~\cite{Armstrong07:Erlang,erlang-web-documentation} and especially the virtual actor model of Orleans~\cite{BykovGKLPT11:Orleans, BernsteinBGKT14:Orleans}, are being used to build soft caching layers designed to scale to millions of actors deployed across hundreds of servers using existing cloud-computing infrastructure. These developments indicate that the actor primitive~\cite{Agha:1986:Actors} is increasingly perceived as a natural fit for the design of these applications~\cite{who-is-using-erlang,uses-of-akka,who-is-using-orleans}. By encapsulating state, providing single-threaded semantics for encapsulated state manipulation, and encouraging an asynchronous function-shipping programming paradigm, actors provide a general, modular and concurrent computational model~\cite{Agha:1986:Actors}. The construct of an actor also allows for reasoning about the scalability, elasticity and failure resilience of these services, which aids in their construction in line with the microservices architecture~\cite{Hasselbring16:microservices}. 

The use of actors brings both opportunities and challenges to the architecture of online data-intensive services. On the one hand, the function shipping programming paradigm leads to high locality as opposed to a data shipping approach, while the asynchronous programming model empowers developers to leverage available parallelism in the ever more complex application logic of these services. 
On the other hand, actors expose thorny state management issues such as durability, global state consistency across actors, and failure management to the application developer, which has led to a call for actor-oriented databases~\cite{BernsteinDKM17:ActorDB}.

The vision for actor-oriented databases has primarily argued for integration of data management features, such as transactions and indexing, in a pluggable fashion into actor runtimes~\cite{BernsteinDKM17:ActorDB}. Recently, however, a complementary approach has explored the potential to integrate actor programming models into relational databases while at the same time not giving up robust database state management features~\cite{0001S18:ReactDB}. The latter work investigated, in particular, how to augment the programmability of stored procedures with actor constructs without compromising high performance on modern hardware. 

In the present paper, we make the broader observation and argument that the idea of such \emph{actor-relational database systems}, which integrate the actor programming model with relational data model and querying under transactional guarantees, opens up a solution space to address the challenges faced in the construction and deployment of interactive data-intensive services today. Instead of viewing the database exclusively as a monolithic storage entity, we argue that actor-relational databases provide developers with the view of a logically distributed runtime, where data and application functions accessing the data are co-located.  Actor-relational database systems couple the modular, asynchronous and concurrent actor programming model with configurable and robust state management features. In so doing, this new class of systems combines both abstract computational \emph{and} data models for the specification and deployment of online data-intensive services. The ultimate goal of this combination is to bring about improved programming productivity, high performance, and efficient use of resources. 

It is, unfortunately, not obvious how to concretely realize these promised benefits of actor-relational databases. In this vision paper, we critically analyze and refine the new abstraction of actor-relational database systems by arguing for its need and desirable design principles. Furthermore, we provide a case study of an interactive smart supermarket application to concretize a candidate feature set based on these principles. In so doing, we observe that this new class of systems provides immense research challenges and opportunities, which the database community is in an advantageous position to explore. 

\minisec{Contributions and Roadmap}
This paper proposes a new vision to integrate actors with relational database systems. Specifically, the paper makes the following contributions:
\begin{enumerate}
 \item We review and classify related state-of-the-art system classes in Section~\ref{sec:related} with respect to three core high-level concepts that are fundamental to actor-relational database systems.   
 \item We argue in Section~\ref{sec:why-actor-db} why the time is ripe \emph{now} for the emergence of of actor-relational database systems by analyzing a number of current trends in the requirements, design and deployment of interactive data-intensive applications.
 \item  Given the aforementioned analysis and review of the state of the art, we propose in Section~\ref{sec:design-principles} an initial set of desirable design principles to help guide the development of future actor-relational database systems. As part of this presentation, we provide an overview of an example of an interactive data-intensive application in the domain of smart supermarkets. The example illustrates at a high level the varied features that an actor-relational database system should support. 
\item To illustrate the potential interplay of the design principles with applications, we drill down on candidate feature sets for each principle in Sections~\ref{sec:actor:construct} to \ref{sec:actor:security}. In order to concretize the design principles and their feature sets, we use the case study as a running example and present application pseudocode. For a subset of the features related to modularity, querying, and transactions, we illustrate the promise of improved performance with asynchrony in actor-relational database systems in Section~\ref{sec:evaluation} by implementing and evaluating the application in \reactdb~\cite{0001S18:ReactDB}.
\end{enumerate}

\section{State of the Art}
\label{sec:related}
\begin{table*}
  \begin{centering}
    \newcolumntype{C}{>{\centering\arraybackslash}X}
    \newcolumntype{W}{>{\centering\arraybackslash} m{8cm}}
    \begin{tabularx}{\linewidth}{|W|C|C|C|}
      \hline
      \textbf{System Classes} & \textbf{Encapsulation} & \textbf{Concurrency and Asynchronicity} & \textbf{Declarative Data Management} \\ \hline
      Object-Oriented Languages and Frameworks & \checkmark & \checkmarkwithcross & \xmark \\ \hline
      Actor Languages and Frameworks& \checkmark & \checkmark & \xmark \\ \hline
      Object-Oriented Distributed Computing Systems& \checkmark & \checkmark & \xmark \\ \hline
      Serverless or Reactive Programming Frameworks& \xmark & \checkmark & \xmark \\ \hline
      Relational DBMS& \xmark & \xmark & \checkmark \\ \hline
      Object-Oriented DBMS& \checkmark & \xmark & \checkmark \\ \hline
      Active DBMS& \checkmarkwithcross & \checkmarkwithcross & \checkmark \\ \hline
      \textbf{Actor-Relational DBMS} & \checkmark & \checkmark & \checkmark \\ \hline
    \end{tabularx}
    \vspace{-2ex}
    \caption{Classification of the state of the art systems and abstractions}
        \vspace{-2ex}
    \label{tab:state-of-the-art}
  \end{centering}
\end{table*}
In this section, we identify and discuss state-of-the-art systems and abstractions that are related to actor-relational database systems. 
To categorize the state of the art,
we analyze existing systems and abstractions with respect to three key concepts that are fundamental to actor-relational database systems, namely: 
(1) \textit{Encapsulation} - The programming model should provide well-defined constructs for isolating state and controlling access to it. 

(2) \textit{Concurrency and Asynchrony} - The programming model should provide well-defined primitives for specifying and reasoning about concurrent and asynchronous computations. 

(3) \textit{Declarative Data Management} - The programming model should provide physical data independence, declarative querying over logical state and failure management.

Table~\ref{tab:state-of-the-art} provides a summarized analysis of the related state-of-the-art system classes with respect to these three concepts. We discuss each such system class in more detail below. 

\minisec{Object-Oriented Languages and Frameworks}
Object-orientation was originally proposed to model discrete event simulations and provide language mechanisms for reusable program components~\cite{Dahl90:OOReflections}. Early object-oriented languages such as Simula~\cite{DahlN66:Simula}, Smalltalk~\cite{Ingalls78:Smalltalk}, CLOS~\cite{0068972:CLOS}, Eiffel~\cite{Meyer91:Eiffel}, C++~\cite{Stroustrup86:C++} and Java~\cite{GoslingJS96:Java} grew in mainstream popularity because they provided general-purpose constructs for concept modeling, leading to better program decomposition and reuse. The notion of objects that encapsulate data and define functions to operate on that data provides an abstract modeling and data hiding abstraction that is intuitive and allows for construction of complex systems by protecting local class-level invariants~\cite{Dahl90:OOReflections}. Originally, all of these languages were conceptualized for writing sequential programs, since concurrent programming was not a target feature. 
Over time, some of these languages added support for concurrent programming either using external libraries or by incorporating language features. In line with this evolution, newer concurrent object-oriented programming languages such as Go~\cite{golang} and Rust~\cite{rustlang} have emerged that are becoming increasingly popular. Despite these advances, these languages and frameworks do not target any declarative data management features. 

\minisec{Actor Languages and Frameworks}
Actors were proposed as a model for concurrent computations centered around message passing semantics~\cite{Agha:1986:Actors}. Actors encapsulate state, provide single-threaded semantics for message handling and state manipulation, and support an asynchronous message shipping programming paradigm. Because of these concepts, actor languages and frameworks provide an elegant mechanism to model concurrent and distributed applications~\cite{BriotGL98:earlyactors,erlang-web-documentation,akka-web-documentation, scala-web-documentation}. However, managing actor lifecycle, handling faults, and ensuring high performance of actor runtimes in a distributed infrastructure complicates their usage, which has led to the appeal of virtual actors~\cite{BykovGKLPT11:Orleans} for transactional middleware~\cite{Bernstein13:actormiddleware}. 

Despite advances in actor runtimes such as virtual actors, actors do not provide any declarative state management features. Applications need to choose either main memory or external storage solutions for actor state, depending on their durability and fault-tolerance requirements. Applications are also forced to account for and handle the failure and consistency models of the underlying storage systems employed. In particular, lack of all-or-nothing atomicity leads to complications in application code to ensure consistency of application state under failure. All of these deficiencies have motivated a call for integrating pluggable data management features, such as transactions and indexing, in an actor runtime that is employed as a soft caching layer~\cite{BernsteinDKM17:ActorDB}.

By contrast, our work outlines a vision for integrating actor programming models into relational databases and not database features into an actor runtime. The latter approach, aimed at enriching the Orleans actor runtime~\cite{BykovGKLPT11:Orleans}, lacks in declarative querying features, as the data model exposed is that of a programming language such as C\#. Moreover, this approach casts databases as storage systems, while our vision enriches the storage layer with actor programming models allowing function shipping instead of data shipping. 

\minisec{Object-Oriented Distributed Computing Systems}
A large class of early object-oriented distributed computing systems were proposed to ease programming under the complexities of distribution~\cite{LiskovCJS87:Argus, LazowskaLAFFV81:Eden, Birman85:Isis,DasguptaLA88:Cloud, ChrysanthisRSV86:Gutenberg, Oakley89:Mercury}. In contrast to actor runtimes, state management under faults with high performance was one of the major design considerations of this class of systems. As a result, state encapsulation as well as concurrency primitives with varied support for asynchrony and state consistency semantics under failures were baked into these systems. However, these systems exposed the physical data model of either the language that the system was built in or a specific programming language that it was tailored for, thus lacking physical data independence. Moreover, the lack of declarative query capabilities hindered the declarative data management capabilities of this class of systems.

\minisec{Serverless Computing and Reactive Programming Frameworks}
Serverless computing architectures are becoming popular for building backend web-services because they relieve the application developer from worrying about infrastructure provisioning and deployment while billing only the consumption of the actual usage of resources~\cite{JonasPVSR17:Serverless}. Functions as a service (FaaS) platforms provide a concrete serverless architecture offering by allowing application-defined compute functions written in various supported languages to run on demand for individual requests or events while guaranteeing scalability~\cite{aws-lambda, google-cloud-functions, microsoft-azure-functions, openwhisk}. Although these offerings provide concurrent and asynchronous execution constructs, their event-driven programming model is geared towards building stateless compute functions. As such, any interactions with stateful entities needs to be handled by the application. Moreover, of late reactive systems have garnered a lot of attention because of their advocacy for a message-driven architecture for building fault-tolerant, elastic and responsive systems~\cite{reactivemanifesto}. As a result, the programming models spanning the serverless computing and reactive programming landscape provide support for concurrent and asynchronous programming, but limited support for state encapsulation and no support for declarative data management. 

\minisec{Relational Database Management Systems (RDBMS)}
RDBMS were designed to support declarative querying of data abstracted using the relational data model~\cite{HellersteinSH07:ArchDB}. In order to shield the application developer from concurrent execution of application programs and hardware failures, ACID transactions became the de facto standard for RDBMS. As a performance optimization, stored procedures were introduced to co-locate a sequence of client queries in the database and reduce data transfer costs~\cite{RoweS87:StoredProcedures}. However, RDBMS lack programming constructs for encapsulation as well as concurrent and asynchronous programming. The latter deficiency is unaffected by the use of database partitioning under the hood in RDBMS for deployment in different hardware infrastructures, e.g., multiple machines or multiple cores~\cite{Kemper:2011:HHO:2004686.2005619,Stonebraker:2007:EAE:1325851.1325981,Diaconu:2013:HSS:2463676.2463710, Johnson:2009:SSS:1516360.1516365, Pandis:2010:DTE:1920841.1920959,Tu:2013:STM:2517349.2522713}.
 
\minisec{Object-oriented Database Management Systems (OODBMS)}
OODBMS focused on addressing the impedance mismatch existing between RDBMS and programming languages~\cite{BancilhonDK92:o2book, AtkinsonBDDMZ89:oo-manifesto}. While persistent programming language runtimes tried to bring database support to popular object-oriented languages such as C++~\cite{AtkinsonB87:dbproglangruntime}, OODBMS such as O2 proposed an object-oriented data model with an embedded declarative query language~\cite{LecluseR89:O2}. OODBMS proposed object-orientation for modularity, data encapsulation, behavior specification, and extensibility. However, objects in OODBMS do not have any notion of a thread of control, i.e, objects are not active entities that can execute logic, but rather they are an abstraction to encapsulate data and to define behavior on this data. This issue is likewise observed in the abstraction of objects in object-oriented programming languages~\cite{akka-objects-vs-actors}. Consequently, despite providing encapsulation and declarative data management features, OODBMS lack constructs for concurrent and asynchronous programming.

\minisec{Active DBMS}
Traditional DBMS, e.g., RDBMS and OODBMS are passive systems where any action happens in the database as a reaction to an explicit request from an application. In order to trigger an action in the database based on a set of rules and without the intervention of an application, active DBMS were proposed~\cite{PatonD99:ActiveDBS, Dayal88:ActiveDB}. Although active DBMS provided declarative data management features, the support for object-orientation and encapsulation was largely limited to a few specific systems~\cite{GehaniJ91:Ode} rather than being a feature of the entire class. In addition, one could argue that a limited form of asynchronous programming capabilities could be found in these systems in the form of rule-based action invocations together with transaction mode coupling. However, these mechanisms lead to obscure and complex reasoning about the interleaving of different events and their corresponding actions. Consequently, it remains in practice a challenge to employ them reliably and correctly for asynchronous execution of arbitrary programs. Moreover, these mechanisms do not provide the notion of a computational primitive or a thread of control, thus lacking an explicit primitive for concurrency in the programming model. 

\section{Why Actor-Relational Database Systems Now?}
\label{sec:why-actor-db}
In this section, we outline the technological and application design trends that act as key enablers for the emergence and future use of actor-relational database systems.

\subsection{Stateful, Latency-Sensitive Web Services}
The last few years have witnessed a rapid growth of stateful, scalable, latency-sensitive web services in various application domains, e.g., online games, mobile and social payment platforms, financial trading systems, and IoT edge analytics~\cite{stateful-application-growth}. Web-application frameworks employing asynchronous and reactive programming, actor runtimes and NoSQL data stores have seen widespread adoption for constructing these interactive web services. Increasingly, application logic has been moved away from the database system into the middle tier, repurposing the database system as a fault-tolerant and consistent storage layer. This migration of logic to the middle tier came not only as a response to concerns regarding scalability and latency, but also to address programming flexibility and development productivity~\cite{stateful-service-architectures, paypal-akka}. The growth of microservices as an architectural pattern has made a case for a modular, scalable, fault-tolerant design of these web services to avoid the pitfalls of a monolithic architecture~\cite{Hasselbring16:microservices}.

With growing diversity of application domains, the classic OLTP workloads of these services are evolving to exhibit increasing computational complexity in application logic that interacts with state. This increase in complexity emanates from various trends in application workloads, e.g., heterogeneous workloads consisting of read-mostly transactions that involve statistical operations over scans of small ranges~\cite{WangJFP17:SSN-ReadMostly}, workloads to model financial computations~\cite{Brook15:Finance}, real-time simulations~\cite{Wen:2009:DTA,White:2007:Games,WSS+10:Brace} and online machine learning methods~\cite{Mnih2015:DeepReinforcement}. The increasing adoption of scalable actor runtimes such as Akka, Erlang and Orleans to deploy these services points towards the attractiveness of actor programming models for designing these applications~\cite{stateful-service-architectures}. Actor programming models bring to these applications many benefits, such as the ability to exploit asynchrony in messaging between actors, locality in accesses to encapsulated state within actors, and mobility in allocation of actors to physical processing elements either in the edge or the data center.  

However, actor runtimes shift the responsibility for correct state management under failures to the programmer, which complicates application development and lowers developer productivity~\cite{BailisFFGHS15:FeralCC, DavisTL17:Node-fuzz}. To address these problems, a call has been made to integrate classical database state management functionality in an actor runtime~\cite{BernsteinDKM17:ActorDB}. The data model in these runtimes is low-level, language-dependent and lacks declarative querying capabilities, thus pushing physical design decisions and optimizations into applications. Actor-relational database systems have the potential to simplify application development and portability without compromising programming flexibility and correctness in this growing space of stateful, latency-sensitive applications. To this end, actor-relational database systems provide robust state management guarantees as well as high-level data model and queries, bringing physical data independence to the actor programming model. In doing so, actor-relational database systems increase the programmability of the database system, which can then be conceived as a language runtime with robust state management features instead of just a monolithic storage abstraction.

\subsection{Modular, Elastic, Available and Heterogeneous Applications}
The popularity of microservices points to the importance of a modular design as a solution to manage application complexity, scalability and failures. A modular design helps in better fault isolation, debugging, and profiling of the individual modules as opposed to a monolithic design, thus increasing programming productivity. In addition, a modular design can improve availability by a fail-soft strategy, where faulty modules and its dependents can be identified and made unavailable instead of an entire application. A modular design also aids in targeted monitoring of the modules and better analysis of impacts, which can bring substantial benefits in load provisioning and resource utilization with workload changes. The latter has a direct impact on supporting elasticity for an application, which has become important today given the 24x7 nature of online web services and the changes in workload they go through.

Modularity is also an important building block in supporting the heterogeneous requirements of current web services. For example, an application might have varying durability needs~\cite{SAP-Adaptive-Server}, where: (1)~the entire data need not be durable, and/or (2)~only executions of certain programs need to be durable. Another case can be made for encrypted data~\cite{TuKMZ13:Monomi}, where the entire data need not be encrypted uniformly and engender the associated overheads. Similarly, a case can be made for concurrency control~\cite{TangE18:CormCC}, where different subsets of data and programs can benefit from different concurrency control structures or isolation levels. Currently, these heterogeneous needs are often explicitly managed by applications by deploying different microservice software components, e.g., using different database systems in a polyglot persistence approach~\cite{Hasselbring16:microservices}. This strategy can significantly increase application complexity and maintenance costs.

By contrast, actor-relational database systems offer a modular, concurrent and fault-tolerant programming model and can thus cater to these application needs in a manner that is both more natural and better integrated with data management functionality. By allowing decomposition across actors and supporting actor heterogeneity, actor-relational database systems can allow application programmers to declare the durability, encryption and concurrency needs of each actor, thus making such actors the islands of homogeneity in the application. At the same time, actor-relational database systems hold promise to provide abstractions to manage cross-actor heterogeneity under a unified database programming model.

\subsection{Increasingly Parallel Hardware}
Over the past two decades, computing power has increased dramatically. Initially, this increase came in the form of higher clock rates for processors; of late, in the form of more processing elements (cores) in a single chip. The cost of these elements has gone down dramatically as well, making them widely affordable. Dropping costs and improving performance have also been witnessed in storage and networking technologies, which has given rise to new challenges for database systems to transition these hardware benefits to applications~\cite{Alonso13:harwaresoftwarestar}. Even though database systems lie at the cross-section of system software (OS) and application programs, they lack abstractions to expose the available physical parallelism using a high-level programming model. 

Actor programming models hold promise to fill this gap, as they possess: (1) well-defined concurrency and asynchronous message passing semantics, and (2) a function-shipping programming paradigm. Actors allow for portable specification of applications in terms of high-level, application-defined concurrent computational elements independent of the actual physical hardware and operating system primitives.~With an actor-based specification, the available control-flow parallelism in application functions can be leveraged from a higher level of abstraction to improve application performance. Additionally, actor-relational database systems can exploit the locality information encoded in actors to better target classic data management optimizations, e.g., for index structure layouts, code generation, and transaction affinity, to increase resource and energy efficiency on modern parallel and distributed hardware.

\subsection{The Rise of Edge Computing}
With increasing penetration and use of IoT and mobile devices, the traditional model of cloud computing is giving way to the edge computing model. In this model, some of the storage and computation is distributed across devices on the edge of the network closer to the client devices, as opposed to storing data and executing computations completely in the cloud. Not only does this model improve availability and response time for the client, it also leads to lower network load and higher resource efficiency. The latter is due to reductions in data movement and utilization of the increasing computational power of edge devices. The model also allows for improved security and privacy, since the entire data and computation is not exposed to the cloud infrastructure. Event-driven architectures and serverless computing have been proposed as programming models for edge computing, but these methodologies are largely targeted at stateless computations.

By contrast, actor-relational database systems bring about the opportunity to provide a concrete solution for programming stateful applications for the edge computing model. In particular, the heterogeneity in such a model can be met by the modular, concurrent and logically distributed construct of an actor coupled with an asynchronous, function shipping paradigm. In addition, robust state-management guarantees in the face of failures enrich the programming model in order to simplify application development and improve programming productivity. 

\subsection{Security Risks}
With increasing pervasiveness of software services, security challenges faced by these services continue to grow dramatically. These challenges include issues pertaining to data integrity, access control, authorization, monitoring and auditing of these services. It is equally important to detect security violations and to mitigate them with minimum possible impact on the service operation. With increasing size of application deployments and complexity, it is imperative that software tools support specification of secure application code and help in static and dynamic verification.

Actor-relational database systems open up new possibilities to re-think the database security model for current and future applications. For example, the traditional security model based on users and roles can be augmented in actor-relational database systems with object-capability security, aiding in monitoring information flow on message passing. Having a modular architecture can also enable auditing of security violations and upon incidents, help in limiting unavailability to only the affected actors and/or functions instead of all actors and all functions.

\section{Design Principles}
\label{sec:design-principles}
In this section, we propose a set of design principles for actor-relational database systems based on the analysis of technological trends in Section~\ref{sec:why-actor-db}.~Furthermore, to illustrate how each principle addresses the needs of many emerging interactive data-intensive applications, we introduce a case study of a simplified future IoT supermarket application for next-generation self-checkout~\cite{AmazonGo,PlanetMoneySelfCheckout}. The application, called \emph{SmartMart} in the remainder, models interactions in a supermarket where carts are equipped with sensors to read the physical cart contents and to trigger checkout operations on a backend service.  After presenting an overview of the design principles in this section, we drill down on how these principles could be concretized into a potential set of features for an actor-relational database system in Sections~\ref{sec:actor:construct} to~\ref{sec:actor:security}. 

\begin{figure*}[!t]
        \begin{minipage}{1.0\linewidth}
          \centerline{\includegraphics[width=0.87\linewidth]{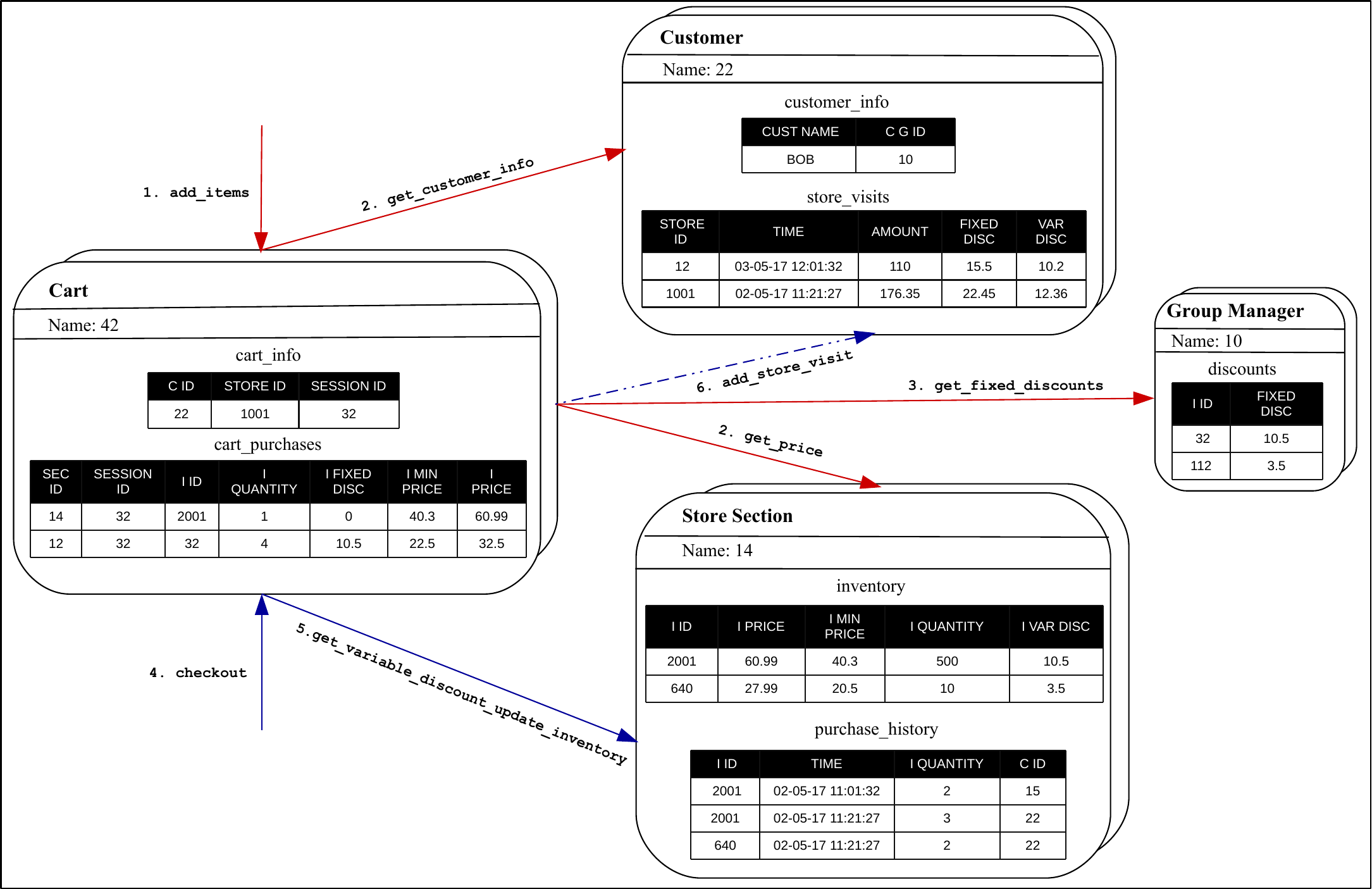}}
        \end{minipage} \hfill
        \caption{Actors in the database of a simplified IoT smart supermarket application. The cart actor supports two transactions, namely \texttt{add\_items} and \texttt{checkout}. Several functions are invoked in response to the \texttt{add\_items} transaction (1). Both \texttt{get\_customer\_info} and \texttt{get\_price} are asynchronously invoked on the customer and multiple store section actors, respectively (2). Once the customer marketing group is obtained, then \texttt{get\_fixed\_discounts} is invoked on the group manager actor (3).
On \texttt{checkout} (4), \texttt{get\_variable\_discount\_update\_inventory} is invoked on each store section asynchronously (5). Finally, a detached transaction \texttt{add\_store\_visit} is invoked for later execution on the customer actor to record the store visit (6).} 
\label{fig:example:app} 
\vspace{-2ex}
\end{figure*}

\subsection{Modularity and Encapsulation using Logical Actors}
In order to tackle growing application complexity, isolate faults, address heterogeneous application requirements, increase programming productivity, and support application elasticity and mobility, modular programming constructs providing encapsulation are required. For high performance, resource efficiency and deployment on varied architectures, it is also desirable that the overhead of modularity be as low as possible and that modules be defined by the application independently of the hardware and software used for deployment. Actors provide this low-overhead \emph{abstract} computational construct in an actor-relational database system. 

Figure~\ref{fig:example:app} shows a set of database actors for the SmartMart application, along with a chain of function invocations triggered by the functionality to \texttt{add\_items} and \texttt{checkout}. In a real application, more operations, such as to remove items from the cart, need to be supported, but these are omitted for brevity. The application is functionally decomposed into actors to represent customers, carts, store sections, and marketing group managers for campaigns. For modularity, \emph{state is encapsulated within each actor}. As a consequence, application developers can isolate logic interacting with different portions of the application state. Moreover, \emph{actors can be deployed in different physical configurations}. For example, in a deployment closer to a traditional web architecture, all actors would run as part of the backend service accessed by gateways at the stores. However, it is also possible to deploy the actors in a way that reflects the natural physical distribution of the application scenario. For example, \texttt{Cart} actors could be deployed in the smart shopping carts, \texttt{StoreSection} actors in store gateways, and the remaining actors in the backend service in the cloud. 

\subsection{Relational Data Model, Declarative Querying and Transactions}
In order to ease the burden of managing application complexity and to increase programming productivity, a well-defined model for concurrency and fault-tolerance is desirable. Single-threaded execution semantics in actor models has been extremely appealing to programmers, hinting at the potential of borrowing and adapting classic database mechanisms such as transactions and more specifically nested transaction models~\cite{WeikumV2002:TxnBigBook}. Moreover, to bring physical data independence to actor models, a high-level data model and declarative query capabilities are required for easy interaction with encapsulated state within the actors, while enabling optimizations for execution of declarative queries.

In Figure~\ref{fig:example:app}, the state of each actor is abstracted by a set of relations and \emph{application functions employ declarative queries against relations}. For example, the customer actor contains relations recording general information about the customers and their store visits, while the cart actor contains relations recording the contents of the cart. The store section actor contains relations recording the inventory for a portion of the store and its purchase history, and the group manager contains relations recording the fixed discount available on each item specialized for a group of customers. In addition, selected \emph{functions perform actions atomically}, in particular the \texttt{add\_items} and \texttt{checkout} functions. When the \texttt{add\_items} function is invoked, we atomically update the cart with the latest prices and fixed discounts. Similarly, when \texttt{checkout} is called, we atomically compute demand-based variable discounts, update inventory, and compute aggregates for the order. This allows for database-style consistency in state manipulation, with isolation under concurrent updates and durability. At the same time, \emph{flexibility in fault tolerance guarantees} can be afforded in some cases. For example, \texttt{checkout} triggers \texttt{add\_store\_visit} as a detached transaction~\cite{PatonD99:ActiveDBS}, which is separately executed at a later time, to record a trace of store visits for the customer. In such a case, independent failure of a customer actor does not preclude the main application functionality from being executed. Similarly, \texttt{add\_items} might not need durability depending on application semantics, since the resulting state can be recomputed from the cart's physical contents albeit with different prices and fixed discounts (if inventory or group manager prices and discounts change).  

\subsection{Communication using Asynchronous Function Shipping}
For latency-sensitive applications to leverage the benefits of increasingly parallel hardware, asynchrony in communication among actors becomes necessary.  By using asynchronous function shipping to communicate with an actor and by utilizing nested invocations of such asynchronous functions, an application can leverage the available application control-flow parallelism to minimize latency, increase locality of data accesses in functions, and thus improve application performance using high-level programming abstractions.

\begin{sloppypar}
In SmartMart, \texttt{add\_items} asynchronously invokes \texttt{get\_customer\_info} and \texttt{get\_price} on customer and store section actors, while \texttt{checkout} invokes \texttt{get\_variable\_discount\_update\_inventory}~on~store section actors.~The asynchrony in invocations exposes \emph{intra-transaction parallelism}, which must interplay cleanly with transactional semantics. Moreover, there are alternative ways in which a programmer could specify this intra-transaction parallelism. On the one hand, the parallelism could arise dynamically from control-flow constructs in the application logic; on the other hand, it could be desirable to carefully consider the semantics of \emph{declarative multi-actor interactions} involving asynchronous function calls. 
\end{sloppypar}

To illustrate why parallelism among actors in an application can be significant depending on application logic, we discuss in detail the computation of discounts in SmartMart. The discount available on an item is classified into two components: (1) fixed discount and (2) variable discount. The fixed discount is customized for a marketing group of customers, while the variable discount is computed based on the demand for the item over a predefined window. Every item has a minimum price as well to ensure that discounts do not overshoot it. The price and fixed discount are computed when items are added to the cart, while the variable discount is only computed at checkout. Since variable discount computation is more expensive, it holds the potential for parallelization in the calls to \texttt{get\_variable\_discount\_update\_inventory}. For an item $i$, if $q_{i,t}$ represents the quantity bought at time $t$ and $S_{i,t}^{k}$ represents the set of quantities from the reverse purchase history of the item starting at $t-1$ and of size at most $k$, then the variable discount is computed using the following formula:

\vspace{-1ex}
\begin{equation} 
vdisc_{i,t} = \frac{q_{i,t}}{pred_{q}(S_{i,t}^{k})} \times VD_{i} 
\label{eq:variable:discount}
\end{equation}

The denominator in the fraction models a target purchase quantity from the history of purchases based on a learning algorithm ($pred_{q}$). $VD_{i}$ is the predefined variable discount that such a predicted target purchase would receive. Thus, the current purchase is scaled by the target purchase and then multiplied with $VD_{i}$ to compute the dynamic variable discount. Such an online computation can add significant latency to an application; however, parallelization of the computation over multiple actors in the database holds promise to bring significant benefits. 

\subsection{Security, Monitoring, and Administration}
In order to address security threats, support for limiting information flow is required from the programming model during design and construction of applications. In addition, monitoring, administration and auditing components provide for further manageability of security violations during execution. By virtue of modularity and encapsulation in actors, a host of language-based security techniques can be provided to enrich the programming model to reason about security of the application during the design phase and to enable software verification. Actor modularity also enables targeted monitoring, auditing, administration and response to security threats in order to minimize availability issues.

For example, in a SmartMart deployment utilizing edge computing resources, the actor-relational database must \emph{enforce security constraints}, such as guaranteeing that \texttt{Cart} actors of a given store only communicate with \texttt{StoreSection} actors in the same store. Moreover, selected actors could be brought offline or restricted in communication with other actors in case a security breach is detected in a part of the deployment.

\section{Modularity and Encapsulation using Logical Actors}
\label{sec:actor:construct}
\subsection*{\manfeature{Logical, Concurrent, Distributed Actors With Location Transparency}{feat:logical:actors}}
\begin{figure}
\centering
\begin{lstlisting}
actor Customer {
 state:   
  relation customer_info (...);

  relation store_visits (...);

 method:
  tuple get_customer_info() {...};

  void add_store_visit(int store_id, timestamp time,
                       float amt, float fixed_disc,
                       float var_disc) {...};
};

actor Group_Manager {
 state:   
  relation discounts (...);

 method:
  list<tuple> get_fixed_discounts(list<int> items) {...};
};
\end{lstlisting}
\vspace{-2ex}
\caption{Actor definitions.}
\label{fig:actor:definition}
\vspace{-3ex}
\end{figure}

\begin{sloppypar}
For modularity, an actor-relational database system provides a programming abstraction of logical actors. A logical actor is an application-defined computational entity encapsulating state that communicates using message passing. Logical actors are concurrent and distributed, since every logical actor is isolated from each other and can execute at the same time independently. A logical actor does not imply a one-to-one mapping to an actual physical element or OS abstraction (e.g., CPU or machine or thread or process) used to implement it, but is merely an application modeling construct.
\end{sloppypar}

Even though logical actors communicate via message passing, we define a logical actor to \emph{not} have a mailbox abstraction as a traditional actor~\cite{Agha:1986:Actors}. We do so to prevent a logical actor from being conceptualized as a low-level interrupt handler where the application developer is forced to implement the message receiving loop. Rather, a logical actor is a reactive entity that processes requests submitted to it from clients or other logical actors. The decision to make logical actors reactive does not preclude the use of application-defined mailboxes, but does not force this design onto every application. 

The application developer can model and structure her application with logical actors as in Figure~\ref{fig:example:app}. Note that the relations displayed in the figure could have been alternatively grouped under a different actor design. Consequently, a given application can be structured in multiple ways depending on how logical actors are defined.

We envision that a logical actor would be understood by developers as a logical thread of control and used to capture the modular units of scaling and available parallelism of the application. For example, the whole set of logical actors in Figure~\ref{fig:example:app} could be instantiated once per store or group of stores.~At finer granularity, the application could be elastically scaled on the number of carts and sections within a store. This abstraction of logical threads of control and parallelism puts actors in actor-relational database systems in contrast to objects in object-oriented database systems~\cite{AtkinsonBDDMZ89:oo-manifesto}. In particular, actors are not simply an abstraction to encapsulate data and define behavior on it. In addition to abstracting computation and parallelism, logical actors enable decomposition of the application in a modular fashion, allowing for better isolation of bugs, containment of failures, and management of runtime application complexity.  

Every logical actor must be uniquely identified by a name. In other words, a logical actor must be assigned a name from a logical namespace by the application. The name of a logical actor acts as its sole immutable identity from a programming perspective. Moreover, names are location independent, which allows for actor mobility. For example in Figure~\ref{fig:example:app}, the name of a customer actor could be chosen to be the implicit primary key of the \texttt{customer\_info} relation, i.e., the customer ID.

\minisec{Logical actors in SmartMart}
Figure~\ref{fig:actor:definition} illustrates through pseudocode actor type definitions for the SmartMart application. The keyword \texttt{actor} is used to specify two application-defined actor types, namely (1)~\texttt{Customer} and (2)~\texttt{Group\_Manager}. An actor type definition must also include the encapsulated state, which is bound to the lifetime of the actor. As expected, actor state is abstracted using relations, whose schema definition is omitted for brevity.\footnote{The full pseudocode for the example is available in Appendix \ref{sec:smartmart:full}. For all code examples, we make simplifications in type annotations and conversions to keep the pseudocode brief.} Each actor type also defines the set of methods that can be invoked on the actor. In contrast to methods in classic object-based models, actor models explicitly define that method calls must be logically shipped for execution on the desired actor, since each actor is logically a concurrent entity. Furthermore, the encapsulated state can only be accessed by invoking methods defined by actors. An actor method can contain any sequence of queries, procedural logic and invocations to other actors. 

\subsection*{\manfeature{Actor Lifetime Management}{feat:actor:names}}
\begin{figure}
  \centering
\begin{lstlisting}
  CREATE ACTORS OF TYPE Cart WITH NAMES IN (42, 43);
  CREATE ACTORS OF TYPE Customer WITH NAMES BETWEEN 22 AND 32;

  DROP ACTORS OF TYPE Cart WITH NAMES IN (42);
\end{lstlisting}
\vspace{-2ex}
\caption{Static Declarative Actor Creation}
\vspace{-3ex}
\label{fig:actor:lifetime:static:declarative}
\end{figure}
Since the programming model exposes primitives for specifying application-defined logical actors, a natural question is whether the application needs to manage the life cycle of these actors. Traditionally, actor models have exposed actor lifetime management to the application. However, an actor-relational database system can either choose to manage actor lifetime itself or delegate it to the application by selecting any of the following mechanisms:

(1)~\emph{Dynamic Actor Creation}: Similarly to dynamic memory allocation, the application can be given the control to create and destroy logical actors dynamically. An attempt to create a duplicate logical actor or to destroy a non-existent logical actor must be signaled by appropriate errors. This policy is similar to the model supported by existing actor language runtimes, such as Akka~\cite{akka-web-documentation} or Erlang~\cite{Armstrong07:Erlang,erlang-web-documentation}. 

(2)~\emph{Static Actor Creation}: An actor-relational database system can also choose to manage actor lifetime itself and therefore not provide primitives for actor creation and deletion. In this mechanism, the actor-relational database system creates the illusion of logical actors to be in perpetual existence to the application. This can be supported by (a)~\textit{automatic actor creation} \cite{BykovGKLPT11:Orleans}, where accessing an actor by name automatically creates that actor; or (b)~\textit{declarative actor creation}, where the application declares the names of the logical actors that should be available for the lifetime of the application.

\minisec{Static declarative actor creation in SmartMart}
In a system supporting static declarative actor creation, commands to create and delete actors should be provided. Figure~\ref{fig:actor:lifetime:static:declarative} shows a sample script to create and delete actors with the appropriate types and names, resulting in a number of actors becoming available for use by the application. Any invocation of a method on an actor not created would raise an error. We use integer values as actor names in the SmartMart application for simplicity; in general, an actor name can be any application-defined string.

The declared actors are available for the lifetime of the application from the point of their creation to their deletion, and the actor-relational database system is responsible for managing the lifecycle of these actors. The use of these commands are for administration only and are disallowed from the method body of the actors as per the static actor creation policy. This allows, for example, the database administrator to exercise control over how to map actors to a particular deployment through additional declarative commands.

\section{Relational Data Model, Declarative Querying and Transactions}
\label{sec:actor:transactions}
\subsection*{\manfeature{Explicit Memory Consistency Model}{feat:txns}}
An actor-relational database system should provide primitives to allow an application to control the isolation level across concurrent computations. Although enforcing serializability of computations on the states of all actors in an actor-relational database would be the simplest isolation level to program with for correctness, evidence suggests that applications can tolerate lower isolation levels with alternative guarantees and mechanisms~\cite{BailisFFGHS15:FeralCC, 0002KBDHKFG15:Homeostasis}. At the same time, and in contrast to microservice architectures~\cite{Hasselbring16:microservices}, actor-relational database systems should not shy away from defining consistency semantics for global state manipulation across multiple~actors. 

Isolation levels and their control can be provided in various ways: (1)~Borrowing traditional database isolation levels and exposing them as annotations in actor computations; (2)~Following a turn-based model similar to Orleans~\cite{BernsteinBGKT14:Orleans}; (3)~Employing application-defined invariant-based isolation guarantees~\cite{Bailis15:InvariantCC, 0002KBDHKFG15:Homeostasis}. In all of these alternatives, isolation in sub-computations of a given computation must be considered carefully. For example, the isolation level of a parent computation could be propagated to child computations, the child could remain independent of the parent, or alternatively the isolation level for the parent and the child should be compatible. 

In addition, the notion of \emph{detached transactions} allows for invocations of sub-computations in a separate transactional context, i.e., the sub-computation does not share the isolation level and atomic commitment requirements of the callee~\cite{PatonD99:ActiveDBS}. The callee can specify when the detached computation should be invoked, e.g., on successful commit of the callee, abort of the callee, or either case.

\begin{figure}[t]
\centering
\begin{lstlisting}
nondurable actor Cart {
state:   
 relation cart_info (c_id int, store_id int,
                     session_id int);

 relation cart_purchases (...);
 ...   
 method:
  float checkout(int c_session_id) {
   SELECT * INTO v_cart FROM cart_info;
   timestamp v_c_time := current_time();

   relation r_v_disc :=  
     APPLY actor<StoreSection>.get_variable_discount_update_inventory[sec_id](c_id, c_time, item_list)
     WITH ARGS
       SELECT S.sec_id, v_cart.c_id, v_c_time AS ctime,
              LIST (SELECT i_id, i_quantity,
                           i_price, i_fixed_disc, 
                           i_min_price
                    FROM cart_purchases
                    WHERE sec_id = S.sec_id
                      AND session_id = S.session_id) AS item_list
       FROM (SELECT DISTINCT sec_id
             FROM cart_purchases
             WHERE session_id = c_session_id) S;

   SELECT SUM(amount) amt, SUM(fixed_disc) fixed_disc,
          SUM(var_disc) var_disc
   FROM r_v_disc;

   DETACH actor<Customer>[v_cart.c_id].add_store_visit(
           v_cart.store_id,v_c_time,amt,fixed_disc,var_disc)
   ON COMMIT EXACTLY ONCE;

   return amt;
 }
 ...
};
\end{lstlisting}
\vspace{-2ex}
\caption{Implementation of \texttt{checkout} in the \texttt{Cart} actor.}
\label{fig:actor:cart:checkout}
\vspace{-3ex}
\end{figure}

\minisec{Serializability in SmartMart}
While multiple memory consistency models are possible, in our example we adopt classic database serializability and propagate this same isolation level across all nested method invocations. This design simplifies the application code in Figures~\ref{fig:actor:cart:checkout} and~\ref{fig:actor:cart:add-items}, since the application developer is insulated from concurrent and asynchronous manipulation of actor state within and across multiple transactions. Furthermore, to maintain well-defined results under asynchrony, any two conflicting sub-computations on the same actor must be ordered by the application code via synchronization using futures; otherwise, the actor database system may abort the entire computation at runtime~\cite{0001S18:ReactDB}. In Figure~\ref{fig:actor:cart:checkout}, a detached transaction on the customer actor is invoked so as to record purchase information on successful commit of~\texttt{checkout}.

\subsection*{\manfeature{Fault-Tolerant Actors With Application-Defined Durability}{feat:faults}}
An actor-relational database system should free the application developer from worrying about partial alteration of actor state by an incomplete computation under failure. Consequently, the system should support the classic notion of recoverability, leading to all-or-nothing atomicity of computations. Still, detached transactions running independently from a calling transaction should be provided to allow for flexibility in fault-tolerance guarantees and performance. For example, the logic of the \texttt{add\_store\_visit} call in Figure~\ref{fig:example:app} is executed as a detached transaction with respect to \texttt{checkout}. To establish a fault-tolerance contract in the call, an exactly-once qualifier can be added to the \texttt{add\_store\_visit} invocation. Different detached transaction invocations may have different qualifiers, e.g., for at-most-once or at-least-once~semantics. 

In addition to recoverability, database systems guarantee durability of committed transactions. By contrast, actor runtimes do not provide any durability guarantees for computations, forcing the application developer to store either parts or the entirety of actor state in an external storage system. To allow flexibility of application design, the programming model of an actor-relational database system should fall in-between the two extremes and allow an application to control actor state durability. The programming model should thus support the notion of durability as a property to avoid conflation with physical deployment as is the case in actor systems. For example in Figure~\ref{fig:example:app}, the application may choose \emph{not} to make the cart actor durable, since it can always be reconstructed if needed by reading the contents in the physical shopping cart itself. Alternatively, durability annotations could be specified per computation and not per actor. In contrast to early systems such as Argus~\cite{Liskov:1988:DistProgArgus}, however, such mechanisms must work in conjunction with a high-level data model and declarative querying.

\minisec{All-or-nothing atomicity and durability in SmartMart}
All-or-nothing atomicity frees the developer from worrying about partial changes to state encapsulated by one or many actors, and thus frees the application logic from implementing failure-handling code as in Figures~\ref{fig:actor:cart:checkout} and~\ref{fig:actor:cart:add-items}. Moreover, application-defined durability per actor allows flexible specification of durability requirements. In the example, the \texttt{Cart} actor is annotated as \texttt{nondurable}, while the other actors are durable by default. So, all state manipulation of committed transactions is durable on the \texttt{Customer}, \texttt{Group\_Manager} and \texttt{Store\_Section} actors only. This allows the application to flag cart state as transient, while not giving up all other data management features in implementing cart operations.    

\subsection*{\manfeature{Relational Data Model and Declarative Querying}{feat:data:model}}
The state of a logical actor should be abstracted by the relational data model to provide for physical data independence. An application developer should have full freedom in defining the schema of a logical actor fitting the application needs, thus allowing schema definitions to vary across logical actor types. The latter frees the application developers from worrying about the physical data layout. The application in Figure~\ref{fig:example:app} shows therefore the use of appropriate schemas in the relational model for the different actor types. The actor-relational database system should also provide a declarative query facility over the state encapsulated in a single actor. This query facility provides ease of programming and allows for reuse of existing database query optimization machinery for performance.

\minisec{Relational state in SmartMart}
In Figure~\ref{fig:actor:cart:checkout}, \texttt{cart\_info} exemplifies a relation schema abstracting portion of the encapsulated state of the actor type \texttt{Cart}. The method \texttt{checkout} interacts with the encapsulated \texttt{cart\_info} and \texttt{cart\_purchases} relations using declarative queries in SQL.

\section{Actor Communication using Asynchronous Function Shipping}
\label{sec:actor:communication}
\subsection*{\manfeature{Asynchronous Operation Support}{feat:async}}
Asynchronous messaging is necessary as a communication mechanism for logical actors to provide a programming construct for the application developer to reason about control and data flow dependencies and explicitly expose parallelism in a computation for performance. Asynchronous messaging across actors can be implemented in various ways, e.g., using traditional actor model message passing~\cite{Agha:1986:Actors} or method invocations on objects returning promises~\cite{BykovGKLPT11:Orleans}, to name a few. Irrespective of implementation mechanism, asynchronous logical actor messaging must allow the callee to synchronize on the result of the communication if desired, i.e., if the communication is a send primitive then the client can choose to wait until the message is successfully received; similarly, if the communication is through a remote procedure call, then the client can choose to wait until its result is propagated back. For example in Figure~\ref{fig:example:app}, the functions \texttt{get\_customer\_info} and \texttt{get\_price} are invoked asynchronously within the \texttt{add\_items} transaction. Similarly, multiple calls to \texttt{get\_variable\_discount\_update\_inventory} are invoked asynchronously within \texttt{checkout}. 

By introducing asynchrony in the messaging mechanism, the illusion of a purely sequential program is broken. Racy computations may now be possible, i.e., in the absence of any concurrent computation and given the same input state, the program execution can produce different result states depending on the order in which asynchronous computations are scheduled on a conflicting data item. Either the programming model must disallow statically the formulation of such programs, or such malformed programs must be rejected at runtime. In addition, the impact of asynchronous messaging on the isolation semantics exposed by the programming model of an actor-relational database system must be clearly defined.

\begin{figure}[t]
\centering
\begin{lstlisting}
nondurable actor Cart {
 state:   
   ...

 method:
  int add_items(list<order> orders, int o_c_id) {
    ... // Organize the item ids in orders by store section
      
    map<int,future> results;
    for (section_order : orders_by_store_section) {
      future res := actor<Store_Section>[section_order.sec_id].get_price(section_order.item_ids);
      results.add(section_order.sec_id, res);
    }
      
    ... // Compute list of all ids of ordered items

    future c_g_res := 
            actor<Customer>[o_c_id].get_customer_info();
    int c_g = c_g_res.get();
      
    future disc_res := actor<Group_Manager>[c_g].get_fixed_discounts(ordered_items);

    ... // Generate session_id and update cart_info 
      
    list<tuple> discounts = disc_res.get();

    results.value_list().when_all(); 
      
    ... // Iterate over prices and discounts and store in cart_purchases

    return v_session_id;
  }

  ...
};
\end{lstlisting}
 \vspace{-2ex}
\caption{Implementation of \texttt{add\_items} in the \texttt{Cart} actor.}
 \vspace{-3ex}
\label{fig:actor:cart:add-items}
\end{figure}

\minisec{Futures in SmartMart}
To exemplify this feature, we abstract method invocations on actors in SmartMart as asynchronous function calls returning \emph{futures}, which represent the result of the computation. In Figure~\ref{fig:actor:cart:add-items}, we illustrate how method invocations and result synchronization could be operationalized in an application program. The figure shows the pseudocode of the \texttt{add\_items} method in the \texttt{Cart} actor. For simplicity of the pseudocode, we assume that the cart is private for a customer, and the method is only invoked once for all the items ordered in the cart before a checkout is performed. Within the pseudocode of \texttt{add\_items}, further method invocations to other actors are performed. An invocation of a method on an actor must specify the type of the actor within the \textbf{$<>$} brackets and the name of the actor within the \textbf{[]} brackets, followed by the method and its arguments. For example, in the first loop in the program, we invoke the \texttt{get\_price} method on each of the store sections in the item orders. The call is directed to an actor of type \texttt{Store\_Section}, whose name is given by the section ID. The method gets as argument a list of the items across all item orders for the corresponding section ID. As a result, the method call produces a future. All futures are collected in a map for synchronization at a later time. As such, the subsequent logic in \texttt{add\_items} is executed while the asynchronous calls to \texttt{get\_price} are processed.   

\begin{sloppypar}
The pseudocode employs an imperative style of invocation of asynchronous function calls in different actors, demonstrating the flexibility in actor models in encoding arbitrary control-flow logic and dependencies in application code. For example, after the first loop, an invocation to the customer actor to get the customer group is made, and the future is immediately synchronized upon since the customer group value is necessary for invocation of \texttt{get\_fixed\_discounts} on the group manager actor. So the computation of \texttt{get\_price} may overlap with both \texttt{get\_customer\_info} and \texttt{get\_fixed\_discounts}, while \texttt{get\_customer\_info} is sequential wrt. \texttt{get\_fixed\_discounts}. Furthermore, a method invocation on an actor can trigger further asynchronous method invocations to other actors, thus allowing arbitrary nesting. In such a case, a method invocation only completes when all its nested method invocations complete. Actors support function shipping by design, since methods can only execute on the actor that defines them and hence the notion of locality is implicit. 

Different schemes can be used to synchronize on future results. The calling code can synchronize on the future by invoking \emph{get} when the value is needed. For example, this is done to obtain the result value from the \texttt{get\_fixed\_discounts} method call. Alternatively, multiple futures can be synchronized upon at the same time by calls such as \texttt{when\_all} or \texttt{when\_one} to consume result values when all or any one are available, respectively. In the example, the futures from the price lookups are synchronized with barrier semantics (\emph{when\_all}), after which cart purchases are recorded and a session ID value is returned for later use in checkout. 
\end{sloppypar}

\subsection*{\manfeature{Declarative Multi-Actor Interactions}{feat:multi:actor:query}}
Although asynchronous invocation of functions on actors using futures enables sophisticated embedding of synchronization patterns, it also leads to erosion of declarative program structures. Consequently, it is desirable that actor-relational database systems also support declarative invocation of functions on one or many actors where the result is consumed in a lazy fashion when needed, in line with the asynchronous nature of actor communication. Such a mechanism for declarative interactions among among multiple actors opens up the space for the optimizer to decide on the actual invocation and synchronization pattern that it sees optimal for performance. Although this feature seems to subsume the low-level actor communication mechanism using futures, multi-paradigmatic actor communication constructs are arguably desirable so that the programmer can be in control of using the best communication primitive for her application without solely depending on the optimizer.

\begin{sloppypar}
\minisec{Multi-Actor Method Invocations in SmartMart}~Figure~\ref{fig:actor:cart:checkout} also exemplifies declarative multi-actor interaction through a bulk invocation of \texttt{get\_variable\_discount\_update\_inventory} in all store sections that have participated in the given session.~This invocation uses the construct \texttt{APPLY <ActorType.methodname> WITH ARGS <subquery>}. Similar to a classic bulk \texttt{INSERT} statement syntax in SQL, \texttt{APPLY} defines a head call that operates on attribute names corresponding to the schema defined in the subquery, and evaluates the head call once per tuple returned by the subquery. Note that the list of items passed as input to each invocation is constructed by converting the relational result of a nested query to a list by the cast \texttt{LIST}. Unlike classic \texttt{INSERT} syntax, however, the invoked functions can return values, which are collected into a result relation. In particular, the result of the invocations in the example is a relation with price and discount information per store section. This result relation can be used further in the program. A subsequent SQL query operates on the relation to compute the total amount bought along with total fixed and variable discounts for the checkout. 

A number of issues need to be considered when specifying the semantics of multi-actor interactions using such an \texttt{APPLY} construct.~First, the relation returned by \texttt{APPLY} is not necessarily required to be eagerly constructed.~The~calls~to \texttt{get\_variable\_discount\_update\_inventory}, for example, can have lazy semantics where actual values are only constructed when when they are consumed. This semantics allows the evaluation of \texttt{APPLY} to be inlined with its first result reference, which occurs in the example in the aggregation query computing total amounts and discounts. Second, the query optimizer can order invocation and future evaluation in the \texttt{APPLY} calls for optimal synchronization patterns. In the example, this reordering would be safe, since the resulting relation is not order-dependent due to encapsulation of state in different actors. The latter is true even though \texttt{get\_variable\_discount\_update\_inventory} updates actor state. In general, the safety of invocations over multiple actors, both with declarative and imperative constructs, needs to be carefully considered and supported by, e.g., static and dynamic checks.   
\end{sloppypar}

\section{Security, Monitoring, and Administration}
\label{sec:actor:security}
\begin{figure}[t]
\begin{lstlisting}
REVOKE ACCESS TO ACTORS OF TYPE ALL FROM ACTORS OF TYPE ALL;
  
GRANT ACTORS OF TYPE Cart WITH METHODS IN (add_items)
 ACCESS TO ACTORS OF TYPE Store_Section 
     WITH METHODS IN (get_price)
 AND ACCESS TO ACTORS OF TYPE Customer 
     WITH METHODS IN (get_customer_info)
 AND ACCESS TO ACTORS OF TYPE Group_Manager 
     WITH METHODS IN (get_fixed_discounts);

GRANT ACTORS OF TYPE Cart WITH METHODS IN (checkout)
 ACCESS TO ACTORS OF TYPE Store_Section 
     WITH METHODS IN 
            (get_variable_discount_update_inventory)
 AND ACCESS TO ACTORS OF TYPE Customer
     WITH METHODS IN (add_store_visit);

GRANT ACTORS OF TYPE Cart WITH NAMES IN (12,13,14)
 ACCESS TO ACTORS OF TYPE Store_Section
     WITH NAMES IN (100, 200);  
 \end{lstlisting}
 \vspace{-2ex}
 \caption{Actor-oriented access control}
 \vspace{-3ex}
 \label{fig:security:access:control}
 \end{figure}

\subsection*{\manfeature{Actor-Oriented Access Control}{feat:access:control}}
Actor encapsulation and modularity provide security by ensuring that an actor's state can only be accessed through its methods, thus localizing security breaches. This enables standard static verification of illegal accesses by information flow analysis. However, this mechanism can be enriched further with access control features found in classic RDBMS, in particular fine-grained access control models~\cite{RjaibiB04:LBAC,KabraRS07:PredicatedGrants}. Such an integration would enable, for example, rich access specifications by methods of actor types and/or particular actor names. By allowing both static and dynamic configuration of access control, static verification and debugging utilities can enrich the design process while dynamic checks protect against violations at runtime. In addition, to further protect against unauthorized access, actors should allow specification of encrypted relations in their state and annotation of methods as encrypted to ensure all communication to and from the methods are encrypted as well. 

\begin{sloppypar}
\minisec{Access control in SmartMart}
In the SmartMart example, suppose we wish to configure minimum levels of access to mitigate security risks so that: (Rule~1)~\texttt{add\_items} in \texttt{Cart} actors has access to \texttt{get\_price} in \texttt{Store\_Section} actors, \texttt{get\_fixed\_discounts} in \texttt{Group\_Manager} actors and \texttt{get\_customer\_info} in \texttt{Customer} actors; (Rule~2) \texttt{checkout} in \texttt{Cart} actors has access to \texttt{get\_variable\_discount\_update\_inventory} in \texttt{Store\_Section} actors and \texttt{add\_store\_visit} in \texttt{Customer} actors; and (Rule~3) A set of specific cart instances (e.g., carts 12, 13, and 14) can only interact with the sections of their corresponding physical store (e.g., store sections 100 and 200).
\end{sloppypar}
 
This access configuration can be enforced in an actor-relational database system by the commands shown in Figure~\ref{fig:security:access:control}. The first \emph{REVOKE} statement revokes access rights of all actors to each other. The next \emph{GRANT} statement configures Rule~1 for access privileges of \texttt{add\_items}, while the next statement configures Rule~2 for access privileges of \texttt{checkout} in the \texttt{Cart} actor type. The final statement additionally sets up a finer-grained rule by actor name to enforce Rule~3. Taken as a whole, the set of rules must cleanly compose or otherwise be flagged and rejected. The set of configured rules can be used for static verification and debugging of security violations. In addition, modification of these rules during deployment enables dynamic adaptivity of security policies to meet unforeseen security threats, e.g., by revoking rights from selected actors. 

\subsection*{\manfeature{Administration and Monitoring}{feat:adm}}
 An actor-relational database system should also provide an administrative interface for flexible maintenance, allowing addition and removal of actors, changing resources allocated to them and modifying access control specifications. Furthermore, an actor-relational database system must support targeted monitoring of actors by gathering statistics of actor usage as well as audit traces of actor method executions, potential security violations, and anomalous accesses. Taken together, these functions enable administrators to intervene in the system at a fine granularity, e.g., by removing or deactivating specific actors. 

\begin{sloppypar}
\minisec{Manageability in SmartMart} 
An actor-relational database system should provide monitoring of actor usage and resource utilization (e.g., some store sections being more loaded than others), security violations (e.g., a \texttt{Group\_Manager} actor attempting to access a \texttt{Store\_Section} actor), and audit traces (e.g., traces of \texttt{checkout} and nested method invocations). Administrative support should also enable scaling actors and resources as well as controlling their placement to meet usage fluctuations discovered during monitoring, or changing access control specifications based on security violations, to name a few possibilities.
\end{sloppypar}

\section{Evaluation}
\label{sec:evaluation}

\begin{figure*}[!t]
  \begin{minipage}{0.32\linewidth}
    \centerline{\includegraphics[width=0.95\linewidth]{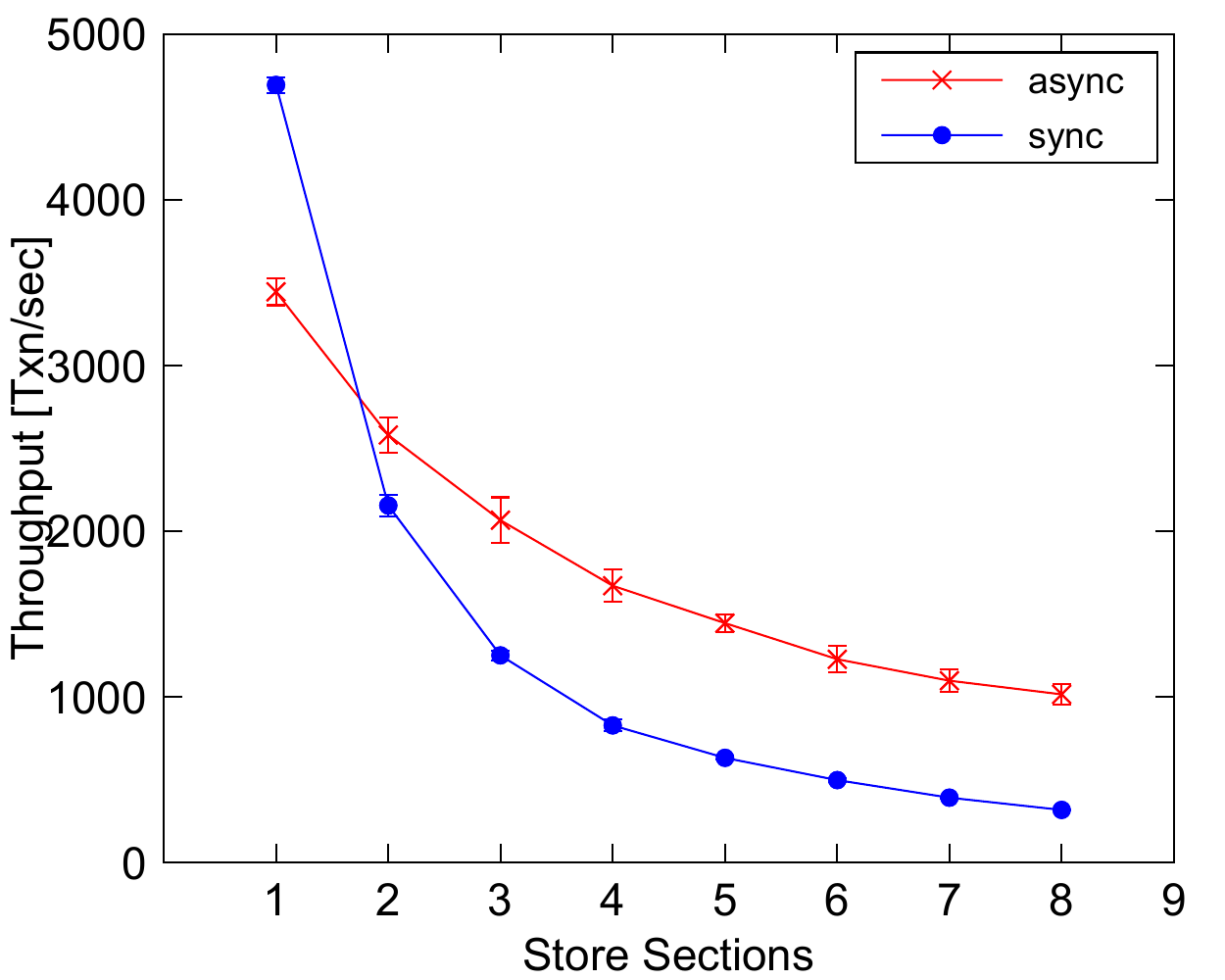}}
    \vspace{-1ex}
    \caption{SmartMart throughput with \\varying order size.} \label{fig:throughput:parallel:scaling}
  \end{minipage} \hfill
    \begin{minipage}{0.32\linewidth}
      \centerline{\includegraphics[width=0.95\linewidth]{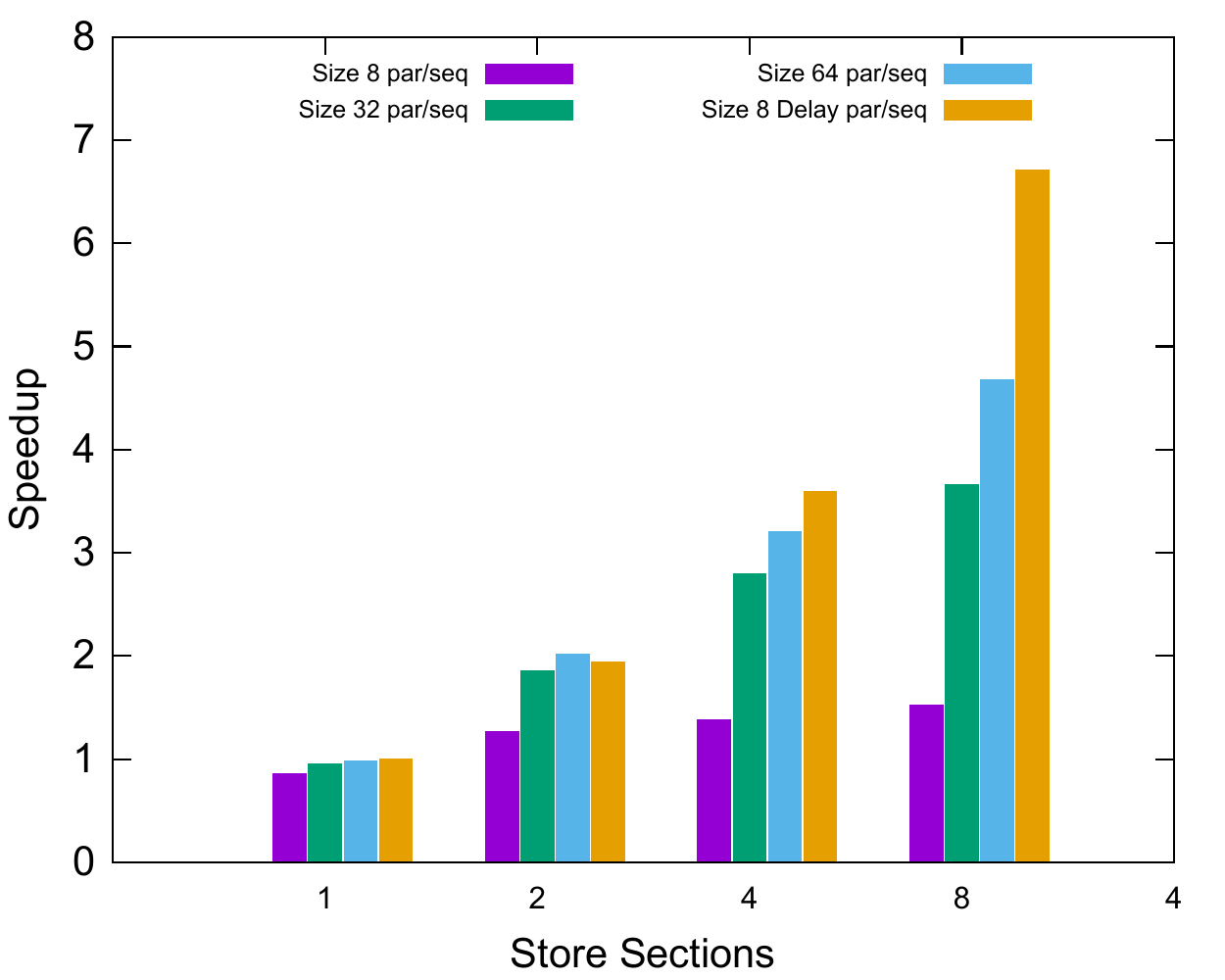}}
      \vspace{-1ex}
      \caption{SmartMart speedup with \\varying parallel work.} \label{fig:throughput:speedup}
      \end{minipage}
  \begin{minipage}{0.32\linewidth}
    \centerline{\includegraphics[width=0.95\linewidth]{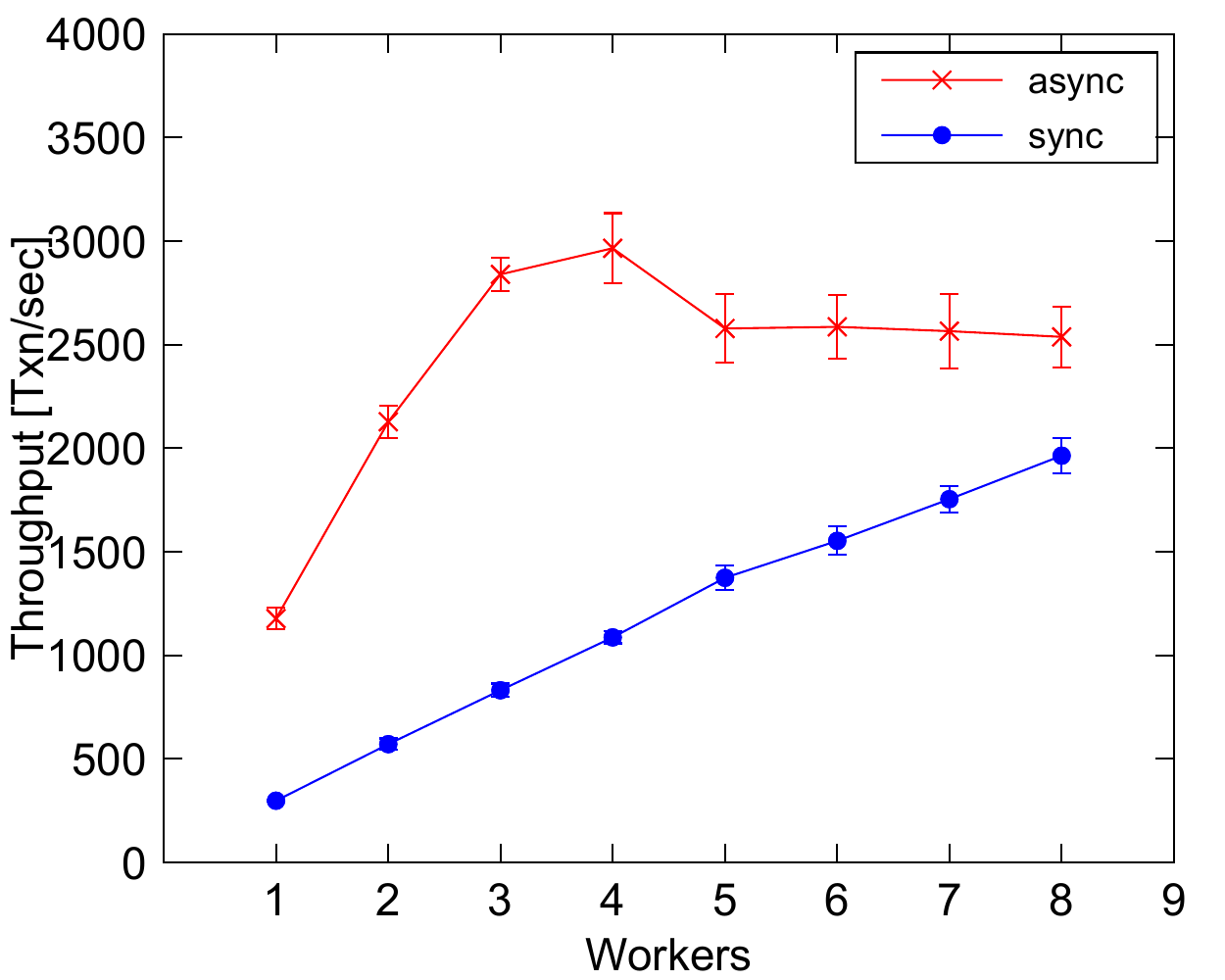}}
    \vspace{-1ex}
    \caption{SmartMart throughput with \\fixed order size and varying workers.} \label{fig:throughput:scalability}
  \end{minipage} \hfill
  \vspace{-1ex}
\end{figure*}

In this section, we evaluate the potential for performance gains from asynchrony in actor-relational databases on modern~hardware.

\subsection{Experimental Setup} 
\label{sec:exp:setup}
\minisec{Hardware}
We focus on multi-core hardware and use a machine with two sockets, each with one eight-core 2.6 GHz Intel Xeon E5-2650 v2 processor with two physical threads per core, leading to a total of 32 hardware threads. Each physical core has a private 32~KB L1 data and instruction cache and a private 256~KB L2 cache. All the cores on the same socket share a last-level L3 cache of 20~MB. The machine has 128~GB of RAM, with half the memory attached to each of the two sockets, and runs 64-bit RHEL Linux 3.10.0. 

\minisec{Workload}
We employ the SmartMart application in our experiments.~To simulate the workings of one store, we created eight \texttt{Store\_Section} actors.~For each, we loaded the \texttt{inventory} relation with 10,000 items and the \texttt{purchase\_history} relation with 300 entries per item for a total of 3,000,000 entries, simulating a history of 120 days where 500 customers on average visit the store per day and buy 50 items each. We fix the number of \texttt{Group\_Manager} actors to 10 and vary the number of \texttt{Cart} actors depending on the experiment. The number of \texttt{Customer} actors is set to 30 times the number of carts. To calculate the variable discount, we used a statistical function based on the mean and variance over the history of purchases as our prediction function, namely: 
\begin{equation} 
pred_{q}(S_{i,t}^{k}) = \mu (S_{i,t}^{k}) \quad + \quad \sigma (S_{i,t}^{k}) 
\end{equation}
We tuned the window size of purchases to be used in the prediction function to correspond to 150 records in the \texttt{purchase\_history} relation, thus calculating a target purchase quantity over 60 days (Equation~\ref{eq:variable:discount}). The entire size allocated after loading was $\sim$3~GB.

\minisec{System Prototype - \reactdb}
We run our experiments in \reactdb, an actor-relational database system prototype with implementations of Features~\ref{feat:logical:actors},~\ref{feat:actor:names},~\ref{feat:txns}, and~\ref{feat:async}~\cite{0001S18:ReactDB}. \reactdb\ provides a relational actor (reactor) programming model that supports the relational data model and querying along with asynchronous procedure calls across actors while guaranteeing globally serializable execution of programs. The system uses a combination of optimistic concurrency control~\cite{Tu:2013:STM:2517349.2522713} and two phase commit protocols. The system also provides low-level control of the database architecture and its deployment by supporting a flexible mapping procedure of reactors to containers and transaction executors.

\begin{sloppypar} 
\minisec{\reactdb\ Deployment Configuration}~We configured the deployment architecture of \reactdb\ in two modes, namely: (1)~\textsf{sync}: all method invocations across actors are executed synchronously and in the same thread to represent sequential execution; and (2)~\textsf{async}: a method invocation on a \textsf{Store\_Section} actor is dispatched to the transaction executor (thread pool) mapped to the actor for execution. Note that our experiments were conducted without any durability of transactions since it is not yet supported by \reactdb. We tuned the thread pool sizes of transaction executors to minimize queuing delays and maximize usage of physical cores.
\end{sloppypar}

We map each \texttt{Cart} actor to a disjoint transaction executor pinned to each of the eight physical cores in the first socket within a single database container. We allocate worker threads such that each worker thread generating method invocations on a \texttt{Cart} actor is mapped to the hyper-threaded core of the corresponding cart to simulate client affinity. For \textsf{async}, all actors except of type \texttt{Store\_Section} that are involved in a cart transaction are mapped to the same database container as the cart actors; they are also mapped to all the transaction executors within the database container to ensure synchronous access within the thread accessing the \texttt{Cart} actor. Invocations to methods of \texttt{Store\_Section} actors, however, are dispatched for asynchronous execution. Each of the \texttt{Store\_Section} actors are mapped to a separate container consisting of a single transaction executor pinned to each of the eight physical cores on the second socket.   

\begin{sloppypar}
\minisec{Methodology}~A worker runs interactions consisting of (1) \texttt{add\_items}, and on its successful commit, (2) \texttt{checkout}. We measure the average latency and throughput of the entire interaction using an epoch-based measurement strategy~\cite{DBLP:journals/pvldb/DifallahPCC13}. Each epoch consists of two~seconds, and we report averages and standard deviations of successful interactions over 20 epochs. Workers choose customer IDs from a uniform distribution. The items and store sections in orders are also chosen from a uniform distribution for a configurable number of store sections and items per store section in the order. 
\end{sloppypar}

\subsection{Can asynchrony and parallelism help cope with increasing work in \textsf{SmartMart}?}
\label{sec:exp:parallel:scaling}
We first study the effect of increasing both work and asynchronicity in method calls from \emph{a single worker}. We vary the number of store sections from 1 to 8 while keeping the number of items ordered from each section fixed at 4, thus varying the size of the order from 4 to 32. Figure~\ref{fig:throughput:parallel:scaling} shows that the throughput of \textsf{sync} degrades with increasing order size given the sequential execution of the methods. The slope of the curve also decreases with store sections since the increase in the order size is constant, and hence has a smaller impact as the order size grows. By contrast, \textsf{async} has lower throughput when the number of store sections is one, but reaches 3.2x higher throughput than \textsf{sync} for 8 store sections. In the beginning, \textsf{async} suffers from dispatch overheads and lack of sufficient asynchrony as opposed to shared memory accesses in \textsf{sync}. However, as the number of store sections increases, asynchronicity benefits arise since the variable discount computations across store sections during \texttt{checkout} and price lookups during \texttt{add\_items} are overlapped to utilize parallel resources. 

\subsection{What speedups can we achieve with increasing parallel resources and parallelizable work in \textsf{SmartMart}?}
\label{sec:exp:speedup}
To characterize the limits of asynchronicity gains we can achieve with the actor-relational database system prototype, we evaluate the speedup obtained for the \textsf{async} deployment compared to the \textsf{sync} deployment of the SmartMart benchmark as we vary the available parallel resources. We kept the total number of items in an order fixed and varied the number of store sections that those items are ordered from. For example, when the total number of items is $N$ and the total number of store sections that items are ordered from is $k$, then $N/k$ items are ordered from each store section. 

Figure~\ref{fig:throughput:speedup} shows the throughput speedups for our experiment with three different sizes of N, namely 8, 32, and 64, which are represented by \textsf{Size N par/seq} bars. As an extra control, we created another variant of variable discount computation where we replaced the scan over the history by an artificial delay of~3 msec created by random number generation to simulate complex prediction calculations. Such complex calculations would increase the ratio of parallel to sequential work without increasing the database footprint and the sequential commit cost. The variant is represented by the \textsf{Size 8 Delay par/seq} bar in Figure~\ref{fig:throughput:speedup}. We increased the delay by this specific amount so as to obtain close to 7x speedup, which we pre-calculated based on the ratio of sequential to parallel work in the interactions. Note that we computed the throughput speedup by calculating the ratio of \textsf{async} and \textsf{sync} throughput. 

We can see that at one store section, we have speedups of slightly less that one. This effect arises due to the lack of asynchronous execution and the small overhead of dispatch. As we increase the number of store sections, however, we can see that the speedups obtained increase as well for all transaction sizes. However, the increase is more pronounced in variants where the ratio of parallel to sequential work in the transactions is larger. Note that this effect is what is expected by application program structure, since a non-trivial fraction of the complex interaction logic, including \texttt{add\_items} and parts of \texttt{checkout}, is sequential. 

At the maximum available parallelism of eight store sections, the speedup of \textsf{Size 8 par/seq} is 1.52, the speedup of \textsf{Size 32 par/seq} is 3.66, the speedup of \textsf{Size 64 par/seq} is 4.67, and the speedup of \textsf{Size 8 Delay par/seq} is 6.8. According to Amdahl's law, in order to get a speedup of 7, 5, 4, 3 and 2 with 8 parallel resources, the parallelizable work must be 98\%, 91\%, 86\%, 76\% and 57\% of the entire work, respectively. This points to the potential of asynchronicity gains that can be leveraged in highly parallel hardware when the amount of complex parallelizable patterns in transactional code increases and the overhead can be kept low.

\subsection{Are asynchrony gains affected by load in \textsf{SmartMart}?}
\label{sec:exp:parallel:load}
By gradually increasing the number of concurrent workers, we study the effect of load on the benefits of asynchronicity observed above. We keep the work fixed to an order size of 32, corresponding to an order across 8 store sections and 4 items from each store section. We increase the number of workers, carts and customers in the experiment. Figure~\ref{fig:throughput:scalability} show the throughput observed. While \textsf{sync} exhibits excellent throughput scalability as we increase the number of workers, the throughput of \textsf{async} scales well until three workers and then degrades before roughly stabilizing. This is because at three workers, the \texttt{Store\_Section} actors are close to full resource utilization (CPU cores at~88\%), maxing out at four workers and then becoming the bottleneck. 

Despite the queuing effects, \textsf{async} still outperforms \textsf{sync} because of the amount of physical resources being utilized by it, namely 16 cores with intra-transaction parallelism as opposed to 8 cores in sequential execution. We did not perform measurements for more than 8 workers, since the hardware does not have enough physical cores to sustain our setting for \textsf{async}. Nevertheless, we would expect a crossover with \textsf{sync} as load increases. In short, asynchronicity can bring both throughput and latency benefits over a traditional synchronous strategy when load in the database is light to normal and transactions exhibit parallelism.  

During this experiment, we observed abort rates of $\sim$5-7\% despite the small amount of actual logical contention on items. This happens because the OCC protocol of Silo aborts transactions if the version numbers of nodes scanned change at validation time, caused in our experiments by tree splits due to inserts.

\section{Outlook and Conclusion}
This paper has made the case for actor-relational database systems, an emerging data management approach combining the virtues of actor runtimes and classic databases.  Actor-relational database systems comprise logical actors with asynchronous operations as well as transactional features and declarative querying, providing for modularity, parallelism, fault tolerance, and security. We argue that increasingly the world of interactive data-intensive applications will look more like the scenario depicted in Figure~\ref{fig:example:app}, where a combination of complex logic and data management is the norm. Instead of having database systems be relegated to persistency providers in these applications, our call to the database community is for reimagining database programming models and architectures for this new world, and bring together actors with relational database systems. 

\clearpage
\balance
\bibliographystyle{abbrv}
\bibliography{actor-db}

\begin{thebibliography}{10}

\bibitem{Agha:1986:Actors}
G.~Agha.
\newblock {\em Actors: A Model of Concurrent Computation in Distributed
  Systems}.
\newblock MIT Press, Cambridge, MA, USA, 1986.

\bibitem{akka-web-documentation}
Akka documentation, July 2018.
\newblock \url{http://akka.io/docs/}.

\bibitem{akka-objects-vs-actors}
Why modern systems need a new programming model ({Akka} documentation), July
  2018.
\newblock
  \url{https://doc.akka.io/docs/akka/2.5.14/guide/actors-motivation.html}.

\bibitem{uses-of-akka}
Examples of use-cases for {Akka}, July 2018.
\newblock \url{http://doc.akka.io/docs/akka/2.4/intro/use-cases.html}.

\bibitem{Alonso13:harwaresoftwarestar}
G.~Alonso.
\newblock Hardware killed the software star.
\newblock In {\em Proc. {ICDE}}, pages 1--4, 2013.

\bibitem{AmazonGo}
{Amazon Go}.
\newblock Store concept website, July 2018.
\newblock \url{https://www.amazon.com/b?node=16008589011}.

\bibitem{Armstrong07:Erlang}
J.~Armstrong.
\newblock A history of {Erlang}.
\newblock In {\em Proc. {ACM} {SIGPLAN}}, pages 1--26, 2007.

\bibitem{AtkinsonBDDMZ89:oo-manifesto}
M.~P. Atkinson, F.~Bancilhon, D.~J. DeWitt, K.~R. Dittrich, D.~Maier, and S.~B.
  Zdonik.
\newblock The object-oriented database system manifesto.
\newblock In {\em {DOOD}}, pages 223--240, 1989.

\bibitem{AtkinsonB87:dbproglangruntime}
M.~P. Atkinson and P.~Buneman.
\newblock Types and persistence in database programming languages.
\newblock {\em {ACM} Comput. Surv.}, 19(2):105--190, 1987.

\bibitem{aws-lambda}
{AWS Lambda}, August 2018.
\newblock \url{https://aws.amazon.com/lambda/}.

\bibitem{Bailis15:InvariantCC}
P.~Bailis.
\newblock The case for invariant-based concurrency control.
\newblock In {\em Proc. {CIDR}}, 2015.

\bibitem{BailisFFGHS15:FeralCC}
P.~Bailis, A.~Fekete, M.~J. Franklin, A.~Ghodsi, J.~M. Hellerstein, and
  I.~Stoica.
\newblock Feral concurrency control: An empirical investigation of modern
  application integrity.
\newblock In {\em Proc. {ACM} {SIGMOD}}, pages 1327--1342, 2015.

\bibitem{BancilhonDK92:o2book}
F.~Bancilhon, C.~Delobel, and P.~C. Kanellakis, editors.
\newblock {\em Building an Object-Oriented Database System, The Story of {O2}}.
\newblock Morgan Kaufmann, 1992.

\bibitem{BernsteinBGKT14:Orleans}
P.~Bernstein, S.~Bykov, A.~Geller, G.~Kliot, and J.~Thelin.
\newblock Orleans: Distributed virtual actors for programmability and
  scalability.
\newblock Technical Report MSR-TR-2014-41, Microsoft Research, 2014.

\bibitem{Bernstein13:actormiddleware}
P.~A. Bernstein.
\newblock Transactional middleware reconsidered.
\newblock In {\em Proc. {CIDR}}, 2013.

\bibitem{BernsteinDKM17:ActorDB}
P.~A. Bernstein, M.~Dashti, T.~Kiefer, and D.~Maier.
\newblock Indexing in an actor-oriented database.
\newblock In {\em Proc. {CIDR}}, 2017.

\bibitem{Birman85:Isis}
K.~P. Birman.
\newblock Replication and fault-tolerance in the {ISIS} system.
\newblock In {\em Proc. {SOSP}}, pages 79--86, 1985.

\bibitem{BohmDMPS16:Operational}
A.~B{\"{o}}hm, J.~Dittrich, N.~Mukherjee, I.~Pandis, and R.~Sen.
\newblock Operational analytics data management systems.
\newblock {\em {PVLDB}}, 9(13):1601--1604, 2016.

\bibitem{BriotGL98:earlyactors}
J.~Briot, R.~Guerraoui, and K.~L{\"{o}}hr.
\newblock Concurrency and distribution in object-oriented programming.
\newblock {\em {ACM} Comput. Surv.}, 30(3):291--329, 1998.

\bibitem{Brook15:Finance}
A.~Brook.
\newblock Low-latency distributed applications in finance.
\newblock {\em Commun. {ACM}}, 58(7):42--50, 2015.

\bibitem{BykovGKLPT11:Orleans}
S.~Bykov, A.~Geller, G.~Kliot, J.~R. Larus, R.~Pandya, and J.~Thelin.
\newblock Orleans: cloud computing for everyone.
\newblock In {\em Proc. {ACM} {SOCC}}, 2011.

\bibitem{ChrysanthisRSV86:Gutenberg}
P.~K. Chrysanthis, K.~Ramamritham, D.~W. Stemple, and S.~Vinter.
\newblock The gutenberg operating system kernel.
\newblock In {\em Proc. {FJCC}}, pages 1159--1167, 1986.

\bibitem{Dahl90:OOReflections}
O.~Dahl.
\newblock Object orientation and formal techniques.
\newblock In {\em Proc. VDM}, pages 1--11, 1990.

\bibitem{DahlN66:Simula}
O.~Dahl and K.~Nygaard.
\newblock {SIMULA} - an algol-based simulation language.
\newblock {\em Commun. {ACM}}, 9(9):671--678, 1966.

\bibitem{DasguptaLA88:Cloud}
P.~Dasgupta, R.~J. LeBlanc, and W.~F. Appelbe.
\newblock The clouds distributed operating system.
\newblock In {\em Proc. {ICDCS}}, pages 2--9, 1988.

\bibitem{DavisTL17:Node-fuzz}
J.~Davis, A.~Thekumparampil, and D.~Lee.
\newblock Node.fz: Fuzzing the server-side event-driven architecture.
\newblock In {\em Proc. {ACM} {EuroSys}}, pages 145--160, 2017.

\bibitem{Dayal88:ActiveDB}
U.~Dayal.
\newblock Active database management systems.
\newblock In {\em {JCDKB}}, pages 150--169, 1988.

\bibitem{Diaconu:2013:HSS:2463676.2463710}
C.~Diaconu, C.~Freedman, E.~Ismert, P.-A. Larson, P.~Mittal, R.~Stonecipher,
  N.~Verma, and M.~Zwilling.
\newblock Hekaton: Sql server's memory-optimized oltp engine.
\newblock In {\em {Proc. ACM SIGMOD}}, pages 1243--1254, 2013.

\bibitem{DBLP:journals/pvldb/DifallahPCC13}
D.~E. Difallah, A.~Pavlo, C.~Curino, and P.~Cudr{\'{e}}{-}Mauroux.
\newblock Oltp-bench: An extensible testbed for benchmarking relational
  databases.
\newblock {\em {PVLDB}}, 7(4):277--288, 2013.

\bibitem{erlang-web-documentation}
{Erlang} documentation, July 2018.
\newblock \url{https://www.erlang.org/docs/}.

\bibitem{who-is-using-erlang}
Who uses {Erlang} for product development?, July 2018.
\newblock \url{http://erlang.org/faq/introduction.html}.

\bibitem{GehaniJ91:Ode}
N.~H. Gehani and H.~V. Jagadish.
\newblock Ode as an active database: Constraints and triggers.
\newblock In {\em Proc. {VLDB}}, pages 327--336, 1991.

\bibitem{golang}
{The Go Programming Language}, August 2018.
\newblock \url{https://golang.org/}.

\bibitem{google-cloud-functions}
{Google Cloud Functions}, August 2018.
\newblock \url{https://cloud.google.com/functions/}.

\bibitem{GoslingJS96:Java}
J.~Gosling, W.~N. Joy, and G.~L.~S. Jr.
\newblock {\em The Java Language Specification}.
\newblock Addison-Wesley, 1996.

\bibitem{GuinardTMW11:IoT}
D.~Guinard, V.~Trifa, F.~Mattern, and E.~Wilde.
\newblock From the internet of things to the web of things: Resource-oriented
  architecture and best practices.
\newblock In D.~Uckelmann, M.~Harrison, and F.~Michahelles, editors, {\em
  Architecting the Internet of Things}, pages 97--129. Springer, 2011.

\bibitem{Hasselbring16:microservices}
W.~Hasselbring.
\newblock Microservices for scalability.
\newblock In {\em Proc. {ACM/SPEC} {ICPE}}, pages 133--134, 2016.
\newblock Keynote.

\bibitem{HellersteinSH07:ArchDB}
J.~M. Hellerstein, M.~Stonebraker, and J.~R. Hamilton.
\newblock Architecture of a database system.
\newblock {\em Foundations and Trends in Databases}, 1(2):141--259, 2007.

\bibitem{Ingalls78:Smalltalk}
D.~Ingalls.
\newblock The {Smalltalk}-76 programming system.
\newblock In {\em Proc. {ACM} {POPL}}, pages 9--16, 1978.

\bibitem{Johnson:2009:SSS:1516360.1516365}
R.~Johnson, I.~Pandis, N.~Hardavellas, A.~Ailamaki, and B.~Falsafi.
\newblock {Shore-MT}: A scalable storage manager for the multicore era.
\newblock In {\em {Proc. EDBT}}, pages 24--35, 2009.

\bibitem{JonasPVSR17:Serverless}
E.~Jonas, Q.~Pu, S.~Venkataraman, I.~Stoica, and B.~Recht.
\newblock Occupy the cloud: distributed computing for the 99{\%}.
\newblock In {\em Proc. {ACM SOCC}}, pages 445--451, 2017.

\bibitem{KabraRS07:PredicatedGrants}
G.~Kabra, R.~Ramamurthy, and S.~Sudarshan.
\newblock Fine grained authorization through predicated grants.
\newblock {\em Proc. ICDE}, pages 1174--1183, 2007.

\bibitem{0068972:CLOS}
S.~E. Keene.
\newblock {\em Object-oriented programming in {COMMON} {LISP} - a programmer's
  guide to {CLOS}}.
\newblock Addison-Wesley, 1989.

\bibitem{Kemper:2011:HHO:2004686.2005619}
A.~Kemper and T.~Neumann.
\newblock {HyPer}: A hybrid {OLTP\&OLAP} main memory database system based on
  virtual memory snapshots.
\newblock In {\em {Proc. ICDE}}, pages 195--206, 2011.

\bibitem{LazowskaLAFFV81:Eden}
E.~D. Lazowska, H.~M. Levy, G.~T. Almes, M.~J. Fischer, R.~J. Fowler, and S.~C.
  Vestal.
\newblock The architecture of the eden system.
\newblock In {\em Proc. {SOSP}}, pages 148--159, 1981.

\bibitem{LecluseR89:O2}
C.~L{\'{e}}cluse and P.~Richard.
\newblock The {O2} database programming language.
\newblock In {\em Proc. {VLDB}}, pages 411--422, 1989.

\bibitem{Liskov:1988:DistProgArgus}
B.~Liskov.
\newblock Distributed programming in argus.
\newblock {\em Commun. ACM}, 31(3):300--312, 1988.

\bibitem{LiskovCJS87:Argus}
B.~Liskov, D.~Curtis, P.~Johnson, and R.~Scheifler.
\newblock Implementation of argus.
\newblock In {\em Proc. {SOSP}}, pages 111--122, 1987.

\bibitem{Meyer91:Eiffel}
B.~Meyer.
\newblock {\em Eiffel: The Language}.
\newblock Prentice-Hall, 1991.

\bibitem{microsoft-azure-functions}
{Microsoft Azure Functions}, August 2018.
\newblock \url{https://azure.microsoft.com/en-us/services/functions/}.

\bibitem{Mnih2015:DeepReinforcement}
V.~Mnih, K.~Kavukcuoglu, D.~Silver, A.~A. Rusu, J.~Veness, M.~G. Bellemare,
  A.~Graves, M.~Riedmiller, A.~K. Fidjeland, G.~Ostrovski, S.~Petersen,
  C.~Beattie, A.~Sadik, I.~Antonoglou, H.~King, D.~Kumaran, D.~Wierstra,
  S.~Legg, and D.~Hassabis.
\newblock Human-level control through deep reinforcement learning.
\newblock {\em Nature}, 518(7540):529--533, Feb 2015.
\newblock Letter.

\bibitem{Oakley89:Mercury}
H.~Oakley.
\newblock Mercury: an operating system for medium-grained parallelism.
\newblock {\em Microprocessors and Microsystems - Embedded Hardware Design},
  13(2):97--102, 1989.

\bibitem{openwhisk}
{Open Source Serverless Cloud Platform}, August 2018.
\newblock \url{https://openwhisk.apache.org/}.

\bibitem{who-is-using-orleans}
{Who Is Using Orleans?}, July 2018.
\newblock
  \url{https://dotnet.github.io/orleans/Community/Who-Is-Using-Orleans.html}.

\bibitem{Pandis:2010:DTE:1920841.1920959}
I.~Pandis, R.~Johnson, N.~Hardavellas, and A.~Ailamaki.
\newblock Data-oriented transaction execution.
\newblock {\em {PVLDB}}, 3(1-2):928--939, 2010.

\bibitem{PatonD99:ActiveDBS}
N.~W. Paton and O.~D{\'{\i}}az.
\newblock Active database systems.
\newblock {\em {ACM} Comput. Surv.}, 31(1):63--103, 1999.

\bibitem{paypal-akka}
{How PayPal Scaled To Billions Of Transactions Daily Using Just 8VMs}, August
  2016.
\newblock
  \url{http://highscalability.com/blog/2016/8/15/how-paypal-scaled-to-billions-of-}\\\url{transactions-daily-using-ju.html}.

\bibitem{PlanetMoneySelfCheckout}
{Planet Money Episode 730: Self Checkout}.
\newblock Online podcast, October 2016.
\newblock
  \url{http://www.npr.org/sections/money/2016/10/19/498571623/episode-730-self-checkout}.

\bibitem{reactivemanifesto}
{The Reactive Manifesto}, August 2018.
\newblock \url{https://www.reactivemanifesto.org/}.

\bibitem{RjaibiB04:LBAC}
W.~Rjaibi and P.~Bird.
\newblock A multi-purpose implementation of mandatory access control in
  relational database management systems.
\newblock {\em Proc. VLDB}, pages 1010--1020, 2004.

\bibitem{RoweS87:StoredProcedures}
L.~A. Rowe and M.~Stonebraker.
\newblock The {POSTGRES} data model.
\newblock In {\em Proc. {VLDB}}, pages 83--96, 1987.

\bibitem{0002KBDHKFG15:Homeostasis}
S.~Roy, L.~Kot, G.~Bender, B.~Ding, H.~Hojjat, C.~Koch, N.~Foster, and
  J.~Gehrke.
\newblock The homeostasis protocol: Avoiding transaction coordination through
  program analysis.
\newblock In {\em Proc. {ACM} {SIGMOD}}, pages 1311--1326, 2015.

\bibitem{rustlang}
{The Rust Programming Language}, August 2018.
\newblock \url{https://www.rust-lang.org/en-US/}.

\bibitem{SAP-Adaptive-Server}
{SAP Adaptive Server Enterprise} 16.0, new features summary {SAP Adaptive
  Server Enterprise} 16.0 (version 15.5), in-memory and relaxed-durability
  databases., August 2018.
\newblock
  \url{http://infocenter.sybase.com/help/index.jsp?topic=/com.sybase.infocenter.dc01165.1600/doc/html/jon1254178384853.html}.

\bibitem{scala-web-documentation}
{The Scala Actors API}, July 2018.
\newblock \url{http://docs.scala-lang.org/overviews/core/actors.html}.

\bibitem{0001S18:ReactDB}
V.~Shah and M.~A.~V. Salles.
\newblock Reactors: {A} case for predictable, virtualized actor database
  systems.
\newblock In {\em Proc. {ACM} {SIGMOD}}, pages 259--274, 2018.

\bibitem{stateful-application-growth}
{The Inevitable Rise of the Stateful Web Application}, July 2018.
\newblock \url{https://petabridge.com/blog/stateful-web-applications/}.

\bibitem{stateful-service-architectures}
Making the case for building scalable stateful services in the modern era,
  October 2015.
\newblock
  \url{http://highscalability.com/blog/2015/10/12/making-the-case-for-building-scalable-}\\\url{stateful-services-in-t.html}.

\bibitem{Stonebraker12:NewSQL}
M.~Stonebraker.
\newblock New opportunities for new {SQL}.
\newblock {\em Commun. {ACM}}, 55(11):10--11, 2012.

\bibitem{Stonebraker:2007:EAE:1325851.1325981}
M.~Stonebraker, S.~Madden, D.~J. Abadi, S.~Harizopoulos, N.~Hachem, and
  P.~Helland.
\newblock The end of an architectural era: (it's time for a complete rewrite).
\newblock In {\em {Proc. VLDB}}, pages 1150--1160, 2007.

\bibitem{Stroustrup86:C++}
B.~Stroustrup.
\newblock {\em The {C++} Programming Language, First Edition}.
\newblock Addison-Wesley, 1986.

\bibitem{TangE18:CormCC}
D.~Tang and A.~J. Elmore.
\newblock Toward coordination-free and reconfigurable mixed concurrency
  control.
\newblock In {\em Proc. {USENIX} {ATC}}, pages 809--822, 2018.

\bibitem{TuKMZ13:Monomi}
S.~Tu, M.~F. Kaashoek, S.~Madden, and N.~Zeldovich.
\newblock Processing analytical queries over encrypted data.
\newblock {\em {PVLDB}}, 6(5):289--300, 2013.

\bibitem{Tu:2013:STM:2517349.2522713}
S.~Tu, W.~Zheng, E.~Kohler, B.~Liskov, and S.~Madden.
\newblock Speedy transactions in multicore in-memory databases.
\newblock In {\em {Proc. SOSP}}, pages 18--32, 2013.

\bibitem{WSS+10:Brace}
G.~Wang, M.~A.~V. Salles, B.~Sowell, X.~Wang, T.~Cao, A.~J. Demers, J.~Gehrke,
  and W.~M. White.
\newblock Behavioral simulations in mapreduce.
\newblock {\em PVLDB}, 3(1):952--963, 2010.

\bibitem{WangJFP17:SSN-ReadMostly}
T.~Wang, R.~Johnson, A.~Fekete, and I.~Pandis.
\newblock Efficiently making (almost) any concurrency control mechanism
  serializable.
\newblock {\em {VLDB} J.}, 26(4):537--562, 2017.

\bibitem{WeikumV2002:TxnBigBook}
G.~Weikum and G.~Vossen.
\newblock {\em Transactional Information Systems: Theory, Algorithms, and the
  Practice of Concurrency Control and Recovery}.
\newblock Morgan Kaufmann, 2002.

\bibitem{Wen:2009:DTA}
Y.~Wen.
\newblock {\em Scalability of Dynamic Traffic Assignment}.
\newblock PhD thesis, Cambridge, MA, USA, 2009.
\newblock AAI0821759.

\bibitem{White:2007:Games}
W.~White, A.~Demers, C.~Koch, J.~Gehrke, and R.~Rajagopalan.
\newblock Scaling games to epic proportions.
\newblock In {\em Proc. ACM SIGMOD}, pages 31--42, 2007.

\bibitem{WhiteKGGD07:Games}
W.~M. White, C.~Koch, N.~Gupta, J.~Gehrke, and A.~J. Demers.
\newblock Database research opportunities in computer games.
\newblock {\em {SIGMOD} Record}, 36(3):7--13, 2007.

\end{thebibliography}
\clearpage
\nobalance
\appendix
\section{\textsf{SmartMart} Implementation Details}
\label{sec:smartmart:full}
\begin{figure}[h]
\centering
\begin{lstlisting}
actor Group_Manager {
 state:   
   relation discounts (i_id int, fixed_disc float);
  
 method:
   list<tuple> get_fixed_discounts(list<int> i_ids) {
     return LIST(SELECT * FROM discounts
                 WHERE  i_id IN (TABLE(i_ids)));
   }
;

actor Customer {
 state:
   relation customer_info (cust_name string, c_g_id int);

   relation store_visits (store_id int, time timestamp,
                          amount float, fixed_disc float,
                          var_disc float);

   encrypted relation passwd (enc_passwd string);
   
 method:
   tuple get_customer_info() {
     SELECT * INTO v_info FROM customer_info;
     return v_info;
   }

   void add_store_visit(int store_id, timestamp time,
           float amt, float fixed_disc, float var_disc) {
     INSERT INTO store_visits
     VALUES (store_id, time, amt, fixed_disc, var_disc);
   }

   encrypted bool authenticate (string enc_passwd) {
     SELECT * INTO v_passwd FROM passwd;
     return validate_fn(enc_passwd, v_passwd);
   }
};
\end{lstlisting}
\caption{Implementation of \texttt{Group\_Manager} and \texttt{Customer} actors.}
\label{fig:actor:impl:customer:group-manager}
\end{figure}
In Figures \ref{fig:actor:impl:customer:group-manager}, \ref{fig:actor:impl:store-section} and \ref{fig:actor:impl:cart}, we present the complete pseudocode of the SmartMart application introduced in the main body of the paper. In the pseudocode, we make use of an additional conversion function \texttt{TABLE} to transform a list of values into a relation. The \texttt{Customer} and \texttt{Group\_Manager} actor functionality in Figure~\ref{fig:actor:impl:customer:group-manager} is straightforward and just interacts with the encapsulated state using declarative queries. In the~\texttt{Customer} actor, we introduce an annotation to make a relation, \texttt{passwd}, and method, \texttt{authenticate}, \emph{encrypted} for security.
\begin{figure}[t]
\centering
\begin{lstlisting}
actor Store_Section {
 state:   
  relation inventory (i_id int, i_price float,
                      i_min_price float,
                      i_quantity int, i_var_disc float);

  relation purchase_history (i_id int, time timestamp,
                             i_quantity int, c_id int);
 
 method:
  list<tuple> get_price(list<int> i_ids) {
   return LIST(SELECT i_price, i_min_price
               FROM inventory
               WHERE i_id IN (TABLE(i_ids)));
   }

  tuple get_variable_discount_update_inventory(int c_id,
              timestamp c_time, list<tuple> ord_items) {
   SELECT SUM(
            (CASE
             WHEN i_price > i_fixed_disc + i_var_disc
             THEN i_price - (i_fixed_disc + i_var_disc)
             ELSE i_min_price) * i_quantity) AS amount,
          SUM(i_fixed_disc * i_quantity) AS fixed_disc,
          SUM(
            (CASE
             WHEN i_price > i_fixed_disc + i_var_disc
             THEN i_var_disc
             ELSE (i_price - i_min_price - i_fixed_disc))
                   * i_quantity) AS var_disc
   INTO v_totals
   FROM
     (SELECT ph.i_id, o.i_quantity,
            (o.i_quantity / (ph.i_avg + c * ph.i_stddev))
             * inv.i_var_disc AS i_var_disc,
             o.i_min_price, o.i_price, o.i_fixed_disc             
     FROM (SELECT i_id,
                  AVG(i_quantity)
                      OVER (PARTITION BY i_id
                            ORDER BY time DESC
                            ROWS BETWEEN CURRENT ROW AND K FOLLOWING)
                              AS i_avg,
                  STDDEV(i_quantity)
                      OVER (PARTITION BY i_id
                            ORDER BY time DESC
                            ROWS BETWEEN CURRENT ROW AND K FOLLOWING)
                              AS i_stddev
           FROM purchase_history
           WHERE i_id IN (SELECT i_id
                          FROM TABLE(ord_items)) ph
     INNER JOIN TABLE(ord_items) o ON (o.i_id = ph.i_id)
     INNER JOIN inventory inv ON (inv.i_id = ph.i_id));

   foreach o_i IN ord_items {
     UPDATE inventory
     SET i_quantity = CASE
                      WHEN i_quantity > o_i.i_quantity
                      THEN i_quantity - o_i.i_quantity
                      ELSE 10000
     WHERE i_id = o_i.i_id;
     
     INSERT INTO purchase_history
     VALUES (o_i.i_id, c_time, o_i.i_quantity, c_id);
   }
   return v_totals;
  }
};
\end{lstlisting}
\caption{Implementation of \texttt{Store\_Section} actor.}
\label{fig:actor:impl:store-section}
\end{figure}

\begin{sloppypar}
Figure~\ref{fig:actor:impl:store-section} outlines the implementation of the \texttt{Store\_Section} actor. The \texttt{get\_price} method returns the minimum price and the price of the requested items from the inventory. The \texttt{get\_variable\_discount\_update\_inventory} method first computes a relation (\texttt{ph}) with the mean and standard deviation of purchase quantities for every item in the list of requested items (\texttt{ord\_items}) for a statically defined history window size of \emph{K}. This relation is then joined using an inner join with \texttt{inventory} and the relation representing the ordered items (\texttt{TABLE(ord\_items)}) to get the necessary information required to compute the cumulative price and discounts to be returned for the order. Note that the minimum price has been accounted for in the price and discount computations. Subsequently, for each item in the order, the inventory is updated to reflect the purchase (and replenished if necessary), following which the purchase is recorded in the \texttt{purchase\_history} relation.
\end{sloppypar}
\begin{figure}[t]
\vspace{-2ex}
\centering
\begin{lstlisting}
nondurable actor Cart {
 state:   
  relation cart_info (c_id int, store_id int,
                      session_id int);
   
  relation cart_purchases (sec_id int, session_id int,
                           i_id int, i_fixed_disc float,
                           i_quantity int, i_price float,
                           i_min_price float);
  
 method:
  int add_items(list<order> orders, int o_c_id) {
   // Organize the items ids in orders by store section
   orders_by_store_section = extract_arrange(orders);
   
   map<int,future> results;
   for (section_order : orders_by_store_section) {
    future res := actor<Store_Section>[section_order.
           sec_id].get_price(section_order.item_ids);
    results.add(section_order.sec_id, res);
   }
 
   // Compute list of all ids of ordered items
   ordered_item_ids :=  extract_ids(orders);
   int v_c_g_id = actor<Customer>[o_c_id].
                       get_customer_info().get().c_g_id;
   future disc_res := actor<Group_manager>[v_c_g_id].
                  get_fixed_discounts(ordered_item_ids);
 
   // Generate session_id and update cart_info 
   SELECT session_id + 1 INTO v_s_id FROM cart_info;
   UPDATE cart_info
   SET c_id = o_c_id, session_id = session_id + 1;
   
   list<tuple> discounts := disc_res.get();
   results.value_list().when_all(); 
 
   //Iterate over prices and discounts and store in cart_purchases
   foreach sec_id_res in results {
     foreach i_p in sec_id_res.second.get() {
       fixed_disc := lookup(discounts, i_p.i_id);
       i_quantity := lookup(orders, i_p.i_id);
       INSERT INTO cart_purchases
       VALUES (sec_id_res.first,v_s_id,i_id,fixed_disc,
               i_quantity,i_p.price,i_p.min_price);
     }
   }
   return v_s_id;
 }
 
 float checkout(int c_session_id) {
  SELECT * INTO v_cart FROM cart_info;
  timestamp v_c_time := current_time();
  
  relation r_v_disc :=  
   APPLY actor<StoreSection>.
           get_variable_discount_update_inventory[sec_id]
           (c_id, c_time, item_list)
   WITH ARGS
     SELECT S.sec_id, v_cart.c_id, v_c_time AS ctime,
            LIST (SELECT i_id, i_quantity,i_price,
                         i_fixed_disc, i_min_price
                   FROM cart_purchases
                   WHERE sec_id = S.sec_id
                   AND session_id = S.session_id) AS item_list
     FROM (SELECT DISTINCT sec_id
            FROM cart_purchases
            WHERE session_id = c_session_id) S;
  
  SELECT SUM(amount) amt, SUM(fixed_disc) fixed_disc,
         SUM(var_disc) var_disc
  FROM r_v_disc;
  
  DETACH actor<Customer>[v_cart.c_id].add_store_visit(
             v_cart.store_id,v_c_time,amt,fixed_disc,var_disc)
  ON COMMIT EXACTLY ONCE;
  
  return amt;
 }
};
\end{lstlisting}
\vspace{-3ex}
\caption{Implementation of \texttt{Cart} actor.}
\label{fig:actor:impl:cart}
\end{figure}

\begin{sloppypar}
Figure \ref{fig:actor:impl:cart} shows the implementation of the \texttt{Cart} actor which uses both the imperative and declarative primitives of asynchronous actor communication in \texttt{add\_items} and \texttt{checkout} methods respectively. The \texttt{add\_items} method first constructs a list of item ids by store section from \texttt{orders} provided as input. For brevity in the pseudocode, we represented these data structure interactions as function calls, namely: (1) \texttt{extract\_arrange}, (2) \texttt{extract\_ids}, and (3) \texttt{lookup}. The method then invokes \texttt{get\_price} method calls on each of the \texttt{Store\_Section} actors asynchronously storing the futures in a map data structure for later synchronization. 
\end{sloppypar}

After firing price lookups, the customer group is looked up using a synchronous method call (\texttt{get\_customer\_info()} on the \texttt{Customer} actor. The customer group is then used to invoke \texttt{get\_fixed\_discounts} on the \texttt{Group\_Manager} actor. Finally, synchronization is used to get the discounts and to wait for all price results to become available from store sections. Using imperative constructs, we iterate over the values of the price results (\texttt{results}). For each record in the map data structure, \texttt{.first} and \texttt{.second} are the handle to the key and the value respectively. For each store section ID (\texttt{sec\_id\_res.first}), we invoke \texttt{get()} on the future (\texttt{sec\_id\_res.second}) to get the price values of the items requested from that store section. Note that this call to \texttt{get()} returns immediately, since synchronization on all the futures has been done earlier with \texttt{when\_all()}.
We use the price information in conjunction with lookups in our input \texttt{orders} and fixed discount values (\texttt{discounts}) to then record an entries in \texttt{cart\_purchases} for use during checkout.

The method \texttt{checkout} is explained in Section~\ref{sec:actor:communication}. The function is also included in the definition of the \texttt{Cart} actor in Figure~\ref{fig:actor:impl:cart} for completeness.

\end{document}